\documentclass[preprint2]{aastex}

\usepackage{amsmath}
\usepackage{amssymb}

\usepackage{lscape}

\usepackage{array}
\newcolumntype{I}{r@{}l}
\newcolumntype{L}{>{$}l<{$}}
\newcolumntype{R}{>{$}r<{$}}
\newcolumntype{C}{>{$}c<{$}}
\newcolumntype{E}{>{$}r<{$}@{}>{$}l<{$}}

\shorttitle{Evolution of the Surface of Rotating Massive Stars}
\shortauthors{Heger \& Langer} 

\usepackage{xspace}

\newcommand{\Lsun}{{\ensuremath{\mathrm{L}_{\odot}}}\xspace}
\newcommand{\Msun}{{\ensuremath{\mathrm{M}_{\odot}}}\xspace}
\newcommand{\MsunB}{{\ensuremath{\mathbf{M_{\odot}}}}\xspace}
\newcommand{\Rsun}{{\ensuremath{\mathrm{R}_{\odot}}}\xspace}
\newcommand{\cm}{{\ensuremath{\mathrm{cm}}}\xspace}
\newcommand{\km}{{\ensuremath{\mathrm{km}}}\xspace}
\newcommand{\Sec}{{\ensuremath{\mathrm{s}}}\xspace}

\newcommand{\yr}{{\ensuremath{\mathrm{yr}}}\xspace}
\newcommand{\Myr}{{\ensuremath{\mathrm{Myr}}}\xspace}

\newcommand{\dex}{{\ensuremath{\mathrm{dex}}}\xspace}

\newcommand{\kms}{{\ensuremath{\km\,\Sec^{-1}}}\xspace}

\newcommand{\junit}{{\ensuremath{\cm^2\,\Sec^{-1}}}\xspace}

\newcommand{\Av}[1]{{\left\langle{#1}\right\rangle}}

\newcommand{\isofont}[1]{{\ensuremath{\mathrm{#1}}}\xspace}
\newcommand{\isomass}[1]{{\ensuremath{\isofont{^{#1}}}}\xspace}
\newcommand{\isocharge}[1]{{\ensuremath{\isofont{_{#1}}}}\xspace}
\newcommand{\isotope}[3]{{\ensuremath{\isocharge{#1}\isomass{#2}\isofont{#3}}}\xspace}
\newcommand{\I}[2]{{\ensuremath{\isotope{}{#1}{#2}}}\xspace}
\newcommand{\El}[1]{{\ensuremath{\I{}{#1}}}\xspace}

\newcommand{\Ep}[1]{{\ensuremath{10^{#1}}}\xspace}

\newcommand{\fmu}{{f_{\mathrm{\mu}}}}
\newcommand{\tMS}{{\tau_{\mathrm{MS}}}}
\newcommand{\tRSG}{{\tau_{\mathrm{RSG}}}}
\newcommand{\tBSG}{{\tau_{\mathrm{BSG}}}}
\newcommand{\Xc}{{X_{\mathrm{c}}}}
\newcommand{\nue}{{$\nu_{\mathrm{e}}$}}
\newcommand{\betap}{{$\beta^{+}$}}
\newcommand{\vrot}{{\ensuremath{v_{\mathrm{rot}}}}\xspace}
\newcommand{\fc}{{f_{\mathrm{c}}}} 
\newcommand{\vcrit}{{v_{\mathrm{crit}}}}
\newcommand{\jeq}{{j_{\mathrm{18}}}}
\newcommand{\jeqm}{{j_{\mathrm{18,max}}}}
\newcommand{\vmax}{{v_{\mathrm{max}}}} 

\newcommand{\Rpg}{{(\El{p},$\gamma$)}}
\newcommand{\Rbp}{{(\betap\nue)}}

\newcommand{\lSect}[1]{{\label{sec:#1}}}
\newcommand{\lFig}[1]{{\label{fig:#1}}}
\newcommand{\lTab}[1]{{\label{tab:#1}}}

\newcommand{\pFig}[1]{{\placefigure{fig:#1}}}
\newcommand{\pTab}[1]{{\placetable{tab:#1}}}

\newcommand{\Tabff}[1]{{\ref{tab:#1}}}
\newcommand{\Tab}[1]{{Table~\Tabff{#1}}}

\newcommand{\FigPan}[1]{{\textit{#1}}}
\newcommand{\FIGFF}[2]{{\ref{fig:#2}\FigPan{#1}}}
\newcommand{\Figff}[1]{{\FIGFF{}{#1}}}
\newcommand{\FIG}[2]{{Fig.~\FIGFF{#1}{#2}}}
\newcommand{\Fig}[1]{{\FIG{}{#1}}}

\newcommand{\FIGS}[2]{{Figs.~\FIGFF{#1}{#2}}}
\newcommand{\Figs}[1]{{\FIGS{}{#1}}}

\newcommand{\Sectff}[1]{{\ref{sec:#1}}}

\newcommand{\Sect}[1]{{\S\Sectff{#1}}}

\newcommand{\pan}[1]{{\textit{#1}}}
\newcommand{\Pan}[1]{{Panel~\pan{#1}}}

\newcommand{\PanRange}[2]{{Panels~\pan{#1}--\pan{#2}}}

\begin{document}

\title{
Presupernova Evolution of Rotating Massive Stars II: \\
       Evolution of the Surface Properties}

\author{A. Heger\altaffilmark{1,3} and 
        N. Langer\altaffilmark{2,4}}

\altaffiltext{1}{Astronomy and Astrophysics Department, 
    University of California, Santa Cruz, CA~95064}
\altaffiltext{2}{Astronomical Institute, Utrecht University, 
    Princetonplein~5, NL-3584 CC Utrecht, The Netherlands}
\altaffiltext{3}{Max-Planck-Institut f\"ur Astrophysik, 
    Karl-Schawarzschild-Stra{\ss}e~1, 85740~Garching, Germany}
\altaffiltext{4}{Institut f\"ur Theoretische Physik und Astrophysik, 
    Am Neuen Palais~10, 14469~Potsdam, Germany}

\begin{abstract}
We investigate the evolution of the surface properties of models for
rotating massive stars, i.e., their luminosities, effective
temperatures, surface rotational velocities, and surface abundances of
all isotopes, from the zero age main sequence to the supernova
stage. Our results are based on the grid of stellar models by
\cite{HLW00:I}, which covers solar metallicity stars in the initial
mass range $8-25\,\Msun$. Results are parameterized by initial mass,
initial rotational velocity and major uncertainties in the treatment
of the rotational mixing inside massive stars.

Rotationally induced mixing processes widen the main sequence and
increase the core hydrogen burning lifetime, similar to the effects of
convective overshooting.  It can also significantly increase the
luminosity during and after core hydrogen burning, and strongly
affects the evolution of the effective temperature.  Our models
predict surface rotational velocities for various evolutionary stages,
in particular for blue supergiants, red supergiants, and for the
immediate presupernova stage.

We discuss the changes of the surface abundances due to rotationally
induced mixing for main sequence and post main sequence stars.  We
single out two characteristics by which the effect of rotational
mixing can be distinguished from that of massive close binary mass
transfer, the only alternative process leading to non-standard
chemical surface abundances in massive stars.  A comparison with
observed abundance anomalies in various types of massive stars
supports the concept of rotational mixing in massive stars and
indicates that it is responsible for most of the observed abundance
anomalies.

\end{abstract}



\section{Introduction}

Massive main sequence stars are rapid rotators. Equatorial rotation
velocities span the range $\vrot = 100 - 400\,\kms$, with B~stars
rotating closest to their break-up speed, $\vcrit$
\cite[][]{Fuk82,how97}.

To date, stellar rotation and, in particular, rotationally induced
mixing have not been taken into account in extended studies of
nucleosynthesis in massive stars.  However, it is required to
reproduce, e.g., observed surface abundance anomalies of massive
stars.  In the present paper we predict the temporal evolution of the
surface properties of rotating massive stars.  This is of particular
interest, as it allows a direct comparison of these models with
observations.

Besides abundance changes, lifetime and evolution in the
Hertzsprung-Russell (HR) diagram of rotating stars differ from those
of non-rotating stars.  This is particularly interesting for age
determinations of young stellar clusters.  Since appropriate sets of
rotating stellar models have not been available so far, these studies
have mostly neglected effects of rotation.

The calculations presented here treat rotation in stars according to
the numerical method introduced by \cite{ES76,ES78,pin89}.  Our
implementation is described in more detail in \cite{HLW00:I}.  The
results presented in this paper are based on calculations using the
STERN code which contains a $36$ isotope network including all stable
isotopes up to silicon.  As in \cite{HLW00:I}, we concentrate on Type
II supernova progenitor stars, i.e., stars which evolve sufficiently
massive cores to become core-collapse supernovae, but not massive
enough to experience sufficient mass loss (or mixing) to become
Wolf-Rayet stars (i.e., initial masses in the range
$\sim8-20\,\Msun$).  We also do not consider binaries.

A list of our initial models and the evolution of stellar structure
and angular momentum in the stellar interior are given in
\cite{HLW00:I}.  The discussion of the observable stellar properties,
which is the topic of the present paper, is divided in two major
parts: the main-sequence evolution (\Sect{Hburn}) and the post-main
sequence evolution (\Sect{postH}).  In \Sect{obs} we compare our
results to observations and give our conclusions in \Sect{conclude}.

\section{Central hydrogen burning}
\lSect{Hburn}

Core hydrogen burning comprises about $90\,\%$ of the total stellar
lifetime and is therefore the evolutionary phase which can be most
easily compared with observations.

\subsection{Nucleosynthesis and envelope abundances}
\lSect{HnucEnv}

Central hydrogen burning in massive stars is governed by the CNO cycle
\cite[][]{Bet39,Cla68}.  Equilibrium mass fraction ratios of roughly
$1:50-100:1-5$ for \I{12}C$:$\I{14}N$:$\I{16}O are established in the
core \cite[][]{AM93}.  This deviates distinctly from typical initial
ratios of about $1:0.3:3$ \cite[][]{GN93}.  Most of the initial carbon
and oxygen is converted into \I{14}N.  A subcycle of the CNO process,
which transforms the initial \I{16}O into \I{14}N, passes through
\I{17}F and \I{17}O and is responsible for the production of \I{17}O.
Further characteristic properties of the CNO cycle are the production
of \I{13}C and the destruction of \I{15}N.

In the stellar layers above the convective core the CNO equilibrium
abundances are not reached.  In particular, there is almost no
processing of hydrogen into helium.  However, isotopes which are
produced at lower temperatures than where they are destroyed (e.g.,
\I{13}C and \I{17}O) can have abundance maxima in the envelope.

Other nuclear cycles occurring during hydrogen burning involve, e.g., neon,
sodium, magnesium, and aluminum.  They do not contribute significantly
to the energy generation, but they do alter the abundances of the involved 
isotopes.  As the abundance changes of different isotopes 
occur at different temperatures,  measuring their abundances at the
stellar surface may contribute additional information about
the mixing processes in the stellar interior.

The light isotopes \I3{He}, \I6{Li}, \I7{Li}, \I9{Be}, \I{10}B, and
\I{11}B are destroyed at the low temperature in the stellar
envelope.  This allows to use them to constrain mixing processes in
the upper part of the envelope.  For example, the surface depletion of
lithium is applied to investigate models for the present sun
\cite[][]{TC98}.

\subsection
{Internal mixing: the example of $\mathbf{20\,\MsunB}$ stars}
\lSect{IntHydroMix}

\subsubsection{Mixing and barriers}

\pFig{m-D-E15AB}
\pFig{m-X-20AB}

\Fig{m-X-20AB} compares the internal abundance profiles of the most
abundant isotopes as a function of the Lagrangian mass coordinate,
$m$, in a non-rotating and two rotating $20\,\Msun$ models at core
hydrogen exhaustion.  The flat profiles in the innermost few solar
masses are caused by convection.  Small convective and/or
semiconvective regions \cite[][]{HLW00:I} cause the ``steps'' in the
profile above the convective core.

\Fig{m-X-20AB} also shows the absence of mixing above this region in
the non-rotating case, visible as a kink in the hydrogen, helium, and
oxygen profiles at about $m=10\,\Msun$.  Between $m=10\,\Msun$ and
$m=13\,\Msun$, a sharp decrease in \I{12}C occurs, accompanied by an
increase in \I{14}N.  This is due to partial CNO processing ``in
place'', i.e., without any transport.  Deeper in the core, the carbon
mass fraction increases again and reaches its equilibrium value.
Below $m=10\,\Msun$ \I{14}N approaches its CNO equilibrium value,
while oxygen is depleted.  There is no thermonuclear depletion of
oxygen above $m=10\,\Msun$.  Above $m\approx14\,\Msun$, the CNO
isotopes retain their initial mass fractions.

In contrast, the rotating models show mixing of thermonuclear
processed matter up to the surface of the star.  In Model E20
(\FIG{a}{m-X-20AB}; no inhibiting effects of gradients in the mean
molecular weight ($\mu$-gradients) on the rotationally induced
instabilities) an extended helium gradient reaches from the upper edge
of the convective core up to the surface.  Along with helium, also
nitrogen is enriched while carbon and oxygen are depleted.  Between
$m=5\,\Msun$ and $m=10\,\Msun$, carbon is enhanced.

\lSect{MixHmu}

If $\mu$-gradients are allowed to inhibit the rotationally induced
instabilities (Model E20B; \FIG{b}{m-X-20AB}), the $\mu$-gradient
which forms at the upper edge of the convective core is not smoothed
out, but instead almost completely chokes off any mixing
between core and envelope quite early during the evolution.
Above the ``barrier'' due to the $\mu$-gradient ($\mu$-barrier),
however, mixing is efficient 
(\FIG{b}{m-X-20AB}).  It's stronger than for Model E20, since the
efficiency for compositional mixing was assumed to be $\fc=1/30$ in
Model E20B instead of $\fc=1/100$ for Model E20 \cite[][]{HLW00:I}.
Therefore, a stronger enrichment of nitrogen close to the surface
results in Model E20B, due mainly to the processing of carbon into
nitrogen.  The \I{16}O abundance in the envelope hardly changes in
Model E20B, because the temperature above the $\mu$-barrier is never
high enough to allow for its processing.  Helium, being the
main reason for the $\mu$-barrier, remains small in the envelope.

In \Fig{m-D-E15AB} we show the total diffusion coefficient as
resulting from all mixing processes taken into account at the
beginning, about the middle, and the end of central hydrogen burning
for Models E15 and E15B.  For Model E15B the $\mu$-barrier can be seen
a deep drop in the diffusion coefficient above the convective core at
the two later times, while the $\mu$-barrier is not yet present at the
beginning of core hydrogen burning (solid line).  In Model E15 such a
$\mu$-barrier never forms, since the rotationally induced mixing is
assumed insensitive to $\mu$-gradients, and thus material from the
core can be mixed up to the surface during whole central hydrogen
burning.  However, note that the diffusion coefficient by rotationally
induced mixing processes, dominated by Eddington-Sweet circulation in
the envelope, is lower by a factor three compared to Model E15B, since
$\fc$ is lower by this factor.  Therefore, also the time-scale for
mixing in the envelope is longer by that factor.

\pFig{m-X3-20AB}

\subsubsection{Light element abundances}
\lSect{H-light}

The light elements, which are already destroyed at the rather low
temperatures in the upper part of the envelope, allow to study the
mixing close to the surface.  Lithium is the most fragile and
destroyed first, closely followed by beryllium.  At somewhat higher
temperatures, boron follows, and further in, more than five solar
masses below the surface in the $20\,\Msun$ models, the destruction of
\I3{He} proceeds (\Fig{m-X3-20AB}).

In non-rotating stars, the isotopes are just destroyed ``in place''
wherever the temperature gets high enough, while at lower
temperatures, i.e., closer to the surface, they are not altered
(\Fig{m-X3-20AB}).  Thus their surface abundance during central
hydrogen burning changes only if mass loss is sufficiently strong to
remove the unaltered layers.  This is the case in Model D20 for
lithium and beryllium.  Due to the strong temperature dependence of
the destruction rates, the resulting abundance changes can become
quite dramatic.  Though Model D25 experiences more mass loss than
Model D20, its envelope is more extended and a more massive surface
layer can preserve the initial stellar abundances till the end of
central hydrogen burning -- enough to result in lower depletion of
lithium and beryllium then Model D20.

In rotating stars, however, rotationally induced mixing causes a
downward transport of the light elements.  This has two important
consequences.  First, the light elements are transported into regions
where the temperature is sufficiently high to destroy them.  This
leads to a depletion of these isotopes by the rotating massive stars
during central hydrogen burning.  The faster the mixing in the
envelope, the stronger the depletion.  Secondly, due to mass loss and
expansion of the stars during central hydrogen burning, envelope
layers which have been hot enough to destroy the light elements cool
down, and the rotationally induced mixing can enrich these layers
again with the residual of the light elements in the layers above
before these outer layers get lost by winds.  That is, some of the
light isotopes can actually be \emph{preserved} -- to a small extent
-- by the rotationally induced mixing (e.g., \FIG{a}{m-X3-20AB} for
lithium).  We denote this effect as \emph{rotationally reduced
depletion}.

Obviously, these two effects compete with each other.  As long as mass
loss is not large enough to remove most of the unprocessed matter from
the surface, both the surface abundance and the total mass of the
corresponding isotope will be higher in the non-rotating star.  But if
mass loss is great enough to remove all the unprocessed matter of a
non-rotating star, the rotating model can, for the same amount of mass
loss, actually retain both a higher total mass and surface mass
fraction of the fragile isotope.

\pFig{m-X2-20AB}

In the models of series ``B'', where the mixing in the envelope
is relatively fast, the destruction due to the mixing of the light
elements into hot layers dominates.  This effect increases with
initial angular momentum, and it can destroy lithium, beryllium, and
boron almost completely, and reduce \I3{He} by more than a factor ten,
even at the surface.

\subsubsection{Sodium and beyond}

Nuclear processing of the stable isotopes of neon, sodium, magnesium,
and aluminum occurs only in the hot convective core of the star, i.e.,
below the $\mu$-barrier.  In the core, \I{21}{Ne} and \I{22}{Ne} are
converted to \I{23}{Na} by the reaction
\I{21}{Ne}\Rpg\I{22}{Na}\Rbp\I{22}{Ne}\Rpg\I{23}{Na}.  Thus a
depletion of \I{21}{Ne} and \I{22}{Ne} is accompanied by an enrichment
of \I{23}{Na}.  The long-lived radioactive isotope \I{26}{Al} is
produced from \I{25}{Mg} by proton capture, but decays by electron
capture to \I{26}{Mg} with a lifetime of $\sim1\,\Myr$.  The
processing of \I{26}{Mg} to \I{27}{Al} leads to a slight enrichment of
this aluminum isotope, but it is unimportant for \I{26}{Mg}.  The
abundances of \I{20}{Ne} and \I{24}{Mg} are essentially unaffected.

No sodium or \I{26}{Al} enrichment was found in the envelope of the
non-rotating Model D20, but strong abundance gradients of these
species exist inside the envelope of the rotating Model E20
(inefficient $\mu$-barriers; see \Fig{m-X2-20AB}), and a slight
enrichment is even found at the surface.  In Model E20B the
$\mu$-gradient is weak enough to allow efficient transport of these
processed isotopes from the core into the envelope only during the
early phase of hydrogen burning.  In particular, the rapid conversion
of \I{22}{Ne} to \I{23}{Na} at hydrogen ignition allows it to be
mixed into the envelope before the formation of the $\mu$-barrier.
Due to the more efficient mixing inside the envelope, \I{23}{Na} is
even more enriched at the surface than in Model E20, even though its
total mass inside the envelope is much less.  Similar processes occur
for the other isotopes of the neon/sodium and the magnesium/aluminum
cycles.

The strength of enrichment of nuclear processed matter increases with
initial mass \cite[][]{HLW00:I}.  Several effects contribute: With
increasing stellar mass the stellar interior becomes increasingly
dominated by radiation pressure, and the radiative zone of the star is
closer to the adiabatic stratification, which, for example, increases
the velocity of the Eddington-Sweet circulation \cite[][]{HLW00:I}.
Additionally, the convective core comprises a larger fraction of the
total stellar mass and mass loss increases.  Both help to bring
processed matter to the surface of the star.

\subsection{Evolution in the HR diagram}
\lSect{HburnHRD}

At core hydrogen ignition, a rotating star behaves, in some respect,
as if it had less mass.  It is less luminous \cite[][]{HLW00:I}, i.e.,
burns hydrogen at a lower rate, and ages slower.  This is due to the
reduction of the effective gravity by the centrifugal force
\cite[][]{KMT70}.  At the same time, the radius of a rotating star is
larger than that of a non-rotating star of same mass.  Therefore, it
appears in the HR diagram at lower effective temperature.  This can be
seen in \Figs{MS-HRD-12AB}, \Figff{MS-HRD-15B}, and \Figff{MS-HRD-20B}
by comparing the starting points of the evolutionary track of stars
with different initial rotation rates in the HR diagram
\cite[][]{Bod71,ES76}.

\pFig{MS-HRD-12AB}

\pFig{MS-HRD-15B}

\pFig{MS-HRD-20B}

Without any rotationally induced mixing, rotating stars evolve
essentially parallel to non-rotating stars in the HR diagram during
core hydrogen burning, but at lower luminosity and surface temperature
\cite[][]{KMT70,ES76}.  However, rotationally induced mixing increases
the average mean molecular weight in the star by enriching the
envelope in helium \cite[\Sect{HnucEnv};][]{HLW00:I}.  The degree of
enrichment depends upon the inhibiting effect of $\mu$-gradients and
on initial angular momentum of the star \cite[][]{HLW00:I}.  In
chemically homogeneous stars, $L\propto\mu^4M^3$ \cite[][]{KW91}.
That is, an increasing helium enrichment makes rotating stars to
evolve at higher luminosity which can exceed the luminosity of a
non-rotating star of same initial mass (\Fig{MS-HRD-12AB}).  The
higher luminosity also causes the convective core to recede less or
even to grow \cite[][]{HLW00:I}.

For faster initial rotation, the stars tend to live longer.  At the
beginning of hydrogen burning they are less luminous due to the
reduced effective gravity, and use their fuel supply more sparingly
(\FIG{e1-e3}{t-CNOvL}).  Only at later times they evolve to higher
luminosities because of the higher average mean molecular weight
\cite[][]{HLW00:I}, but at the same time they have a larger fuel supply
and larger convective cores.  This latter effect is analogous to the
effect of ``convective core overshooting'' \cite[e.g.,][]{sch92}.  The helium
enrichment in the envelope can also cause the star to evolve to higher
surface temperatures, and therefore to a position
in the HR diagram where less evolved non-rotating stars of higher
initial mass would be located \cite[][]{Lan92}.

These trends are demonstrated in \Fig{MS-HRD-12AB} using the example
of $12\,\Msun$ stars with different zero-age main sequence (ZAMS)
rotational velocities and for the two cases, rotationally induced
mixing being insensitive (\FIG{a}{MS-HRD-12AB}: Models D12, G12, E12,
and F12) or sensitive (\FIG{b}{MS-HRD-12AB}: Moddels D12, G12B, E12B,
F12B, and H12B) to $\mu$-gradients.  In order to meet the
observational constraints on the surface enrichment, the mixing
efficiency is smaller in the former models \cite[see][]{HLW00:I}.  In
these models, the mixing occurs continuously without any hindrance by
a $\mu$-barrier (\Sect{MixHmu}) and the stars gradually evolve towards
higher luminosities compared to non-rotating stars.  They reach the
end of central hydrogen burning at higher luminosities for higher ZAMS
rotational velocity.

As the fastest rotators remain close to uniform rotation, Model F12
hits the $\Omega$-limit of critical rotation \cite[][]{Lan97:LBV} at
the end of central hydrogen burning.  Therefore, it can not contract
and move back to higher temperatures before the ``turn-off'', like the
other models shown in {\FIG{a}{MS-HRD-12AB}}.  The same occurs for
Model H15B.

In the case where rotationally induced mixing is sensitive to
$\mu$-gradients, the envelope helium enrichment is only weak for the
slowly rotating models ($\vrot\lesssim200\,\kms$), since the
$\mu$-barrier becomes effective early on.  These stars evolve very
similar to the non-rotating models.  In the faster rotating models
(see Models F12B and H12B in \FIG{b}{MS-HRD-12AB}) the rotationally
induced mixing initially operates efficiently enough to penetrate the
growing $\mu$-barrier.  The stars evolve to higher luminosities and
surface temperatures for some time and follow, in the HD diagram, a
track almost perpendicular to the evolution of the non-rotating Model
D12.  During this time, their convective core mass remains about
constant or even grows (in Model H12B).  Eventually, the $\mu$-barrier
becomes strong enough to suppress the rotationally induced mixing, and
the stars continue their evolution in the HR diagram parallel to that
of the non-rotating model, but at a higher luminosity.  This
``$\mu$-turn'' occurs the later the higher the initial rotational
velocity is (for a given initial mass).  It also affects the evolution
of the surface abundances (\Sect{SurfAbu}), because it marks the point
where the chemical evolution of the envelope becomes disconnected from
that of the core.

Very similar behavior is also found for the $15\,\Msun$
(\Fig{MS-HRD-15B}) and $20\,\Msun$ (\Fig{MS-HRD-20B}) stars.  The
effects described above are stronger for the fast-rotating stars of
lower mass because those are closer to their critical rotation rate.
Additionally, the more massive stars are more dominated by radiation
pressure in their interiors, and so helium enrichment affects them
less.

As a result of these dependences of the stellar evolutionary tracks in
the HR~diagram on the initial rotation rate, a given point in the HR
diagram is not uniquely related to a single initial mass, even for
core hydrogen burning stars.  In contrast, stars which differ in
initial mass \emph{and} rotation rate can evolve to the same position
\cite[Fig.~2 of][]{FLV96}.  For example, the fast rotating Models H12B
and H15B share, at the end of their central hydrogen burning, the same
position in the HR~diagram as the slowly rotating models of
$15\,\Msun$ and $20\,\Msun$ (\FIGS{b}{MS-HRD-12AB},
\Figff{MS-HRD-15B}, and \Figff{MS-HRD-20B}).  The rotating models can
also reach positions in the HR diagram to the right of the main
sequence band of non-rotating stars, i.e., the main sequence is
broadened by the action of rotation.

A distinguishing feature, however, between lower-mass fast rotators
and higher-mass slow rotators which share the same position in the HR
diagram, is the different evolutionary time they need to get to that
position.  This can be seen in \FIG{e1-e3}{t-CNOvL}: For example, the
fast rotating Model H12B reaches the same luminosity and also about
the same surface temperature after $25\,\Myr$, while the slowly
rotating $15\,\Msun$ stars terminate central hydrogen burning at the
same position already after only $10\,\Myr$.  Another such
distinguishing feature is the expected surface abundance pattern,
which is discussed in \Sect{SurfAbu}.

Hot stars may become Wolf-Rayet stars with surface hydrogen mass
fractions below $40\,\%$, either due to mass loss or due to mixing
\cite[][]{Mae82,Lan87}.  As mentioned above, stars with more internal
mixing also have more mass loss due to their higher luminosity.  As an
example, the extremely rapidly rotating Model H20B becomes a
Wolf-Rayet star at a central hydrogen abundance of $4\,\%$, i.e.,
during central hydrogen burning (\Fig{MS-HRD-20B}).  Thus,
rotationally induced mixing can significantly influence the lower
limiting initial mass required for Wolf-Rayet star formation during
central hydrogen burning in single stars
\cite[][]{Mae87,Mae99,Lan92,FL95}.

\subsection{Rotational velocities}
\lSect{rotH}

The evolution of the surface rotational velocity for various models is
displayed in \FIG{d1-d3}{t-CNOvL}.  It is affected by several
different contributions: First, the loss of angular momentum due to
stellar winds leads to a slow-down of the star.  For the models shown
in \FIG{d1-d3}{t-CNOvL}, this reduces the initial angular momentum of
the $20\,\Msun$ models by $\lesssim40\,\%$, by $\lesssim20\,\%$ for
the $15\,\Msun$ models, and $\lesssim10\,\%$ for the $12\,\Msun$
models.  The second effect is the increase of the total moment of
inertia of the star, for the stars considered here by about a factor
two from core hydrogen ignition until shortly before core hydrogen
exhaustion, due the expansion of the envelope.  Both the moment of
inertia and the angular momentum of the models are dominated by the
envelope, which stays close to uniform rotation.  Third, the increase
of the stellar radius by about a factor two to three during core
hydrogen burning leads to an increase of the surface rotational
velocity of the stars by the same factor for a given angular velocity,
$\vrot=r\omega$.  This effect almost completely cancels out the
slowing down due to the two preceding effects, as can be seen from the
small change in the surface rotational velocity in
\FIG{d1-d3}{t-CNOvL}.

The rise in rotational velocity at the end of core hydrogen burning is
due to the overall contraction of the star
before ignition of hydrogen shell burning.  The deep
drop of the rotational velocity in Model H20B (\FIG{d3}{t-CNOvL}) is
caused by the angular momentum loss occurring as a consequence of the
strong mass loss as it becomes a Wolf-Rayet star.  Model H12B
(\FIG{d1}{t-CNOvL}) reaches the $\Omega$-limit \cite[][]{Lan97:LBV} at
the end of core hydrogen burning, and therefore strong anisotropic
mass loss is expected in this case \cite[][]{Lan98}.

\subsection{Surface abundances}
\lSect{SurfAbu}

\pFig{t-CNOHe}

\pFig{t-CNOvL}

\pFig{t-BFNaAl}

\pFig{t-BFNaAl}

\pTab{TAMS}

The efficiencies of the rotationally induced mixing in our models have
been calibrated {\cite[][]{HLW00:I}} such that the {\em average} surface
abundance trends for helium and CNO abundances on the main
sequence are consistent with the observations.  However, the
enrichment or depletion of these elements as a function of rotation
rate, time, and initial mass were not adjusted.  Also, the surface
abundances of the rare helium and CNO isotopes, as well as the
abundances of all other elements were not considered in the
calibration process. It is thus essential for a comparison with
observations to also investigate the surface abundances of our models
during the main sequence phase.

\subsubsection{$^3$He, Li, Be, B}

An important feature of rotationally induced mixing is the depletion
of light elements, like lithium and boron, at the surface.  These
elements are destroyed at much lower temperatures than those necessary
for CNO processing.  Therefore, much stronger abundance changes are
induced by rotation.  The rotating main sequence models computed by
\cite{FLV96} predicted that stars which show enrichment in nitrogen
should be strongly depleted in boron, which was found in good
agreement with the observational data by \cite{VLL96}.  This finding
is supported by the present work (\FIGS{a1-a3}{t-CNOHe} and
\FIGFF{a1-a3}{t-BFNaAl}): the boron depletion is much stronger and
appears earlier than the nitrogen enrichment.  Even the slow rotators
with ZAMS rotational velocities of $\sim100\,\kms$, which barely show
any nitrogen enrichment, become depleted in boron by $\sim0.5\,\dex$.

At the end of central hydrogen burning, the non-rotating Model D20
shows a lithium depletion which is more than two orders of magnitude
greater than that of the slowly rotating Model G20.  This is due to
the rotationally reduced depletion (\Sect{H-light}).  In a similar
way, Model E20, where the efficiency of the mixing in the envelope is
less than in Model E20B, experiences less destruction of lithium than
Model D20 (\FIG{a}{m-X3-20AB}).  But the destruction is also less
than in Model E20B, where the efficient mixing would lead to a strong
depletion even without mass loss (\FIG{b}{m-X3-20AB}; see also
{\Fig{m-X3-12AB}} for a model with less mass loss).

While beryllium is preserved at the surface in most of our
non-rotating models, except for Model D20, already our slowest
rotating models, Models G12B, G15B, and G20B, show considerable
destruction of beryllium at the end of central hydrogen burning.
Similar to lithium, those models where the inhibiting effect of the
$\mu$-gradients on rotationally induced mixing are taken into account
but a more efficient chemical mixing is assumed, show much stronger
depletion.  Due to the stronger mass loss, the depletion increases
with increasing initial stellar mass, exceptions being, again,
Model D20, and Models H12B, H15B, and H20B which show an opposite trend,
but at very high depletion factors (\Tab{TAMS}).

The \I3{He} isotope is much less fragile than the species discussed
above and is not depleted at the surface in any of the non-rotating
models and only slightly depleted in the slow to average speed
rotators ($\lesssim200\,\kms$).  In the fast rotators, we find a
depletion of up to almost a factor 100 (Model H20B; \Tab{TAMS}).  The
magnitude of the depletion depends more strongly on the initial
rotation than on the initial mass.

\subsubsection{C, N, O}

The evolution of the surface abundances of \I{12}C, \I{14}N, and
\I{16}O relative to their initial values, and that of the \I4{He} mass
fraction during central hydrogen burning are shown in \Fig{t-CNOHe}
for $12\,\Msun$, $15\,\Msun$ and $20\,\Msun$ stars with different
initial angular momenta.  \Fig{t-CNOvL} displays the corresponding
ratios of elemental carbon, nitrogen and oxygen abundance relative to
their initial values.  The surface abundances at core
hydrogen exhaustion are shown in \Tab{TAMS} for elements up to sodium.

The dominant isotope in the CNO cycle, \I{14}N, is enriched at the
surface at the expense of \I{12}C and \I{16}O.  This enrichment is
very small for the models with a ZAMS equatorial rotational velocity
of $\sim100\,\kms$ ($\lesssim0.1\,\dex$), but for models with typical
rotational velocity ($\sim200\,\kms$) the nitrogen enrichment at the
end of central hydrogen burning is a factor two to three, and it is
increasing with initial mass.  At higher initial rotation rates the
final enrichment factor is about $6$ and $10$ for the models with
$\sim300\,\kms$ and $\sim400\,\kms$ ZAMS rotational velocity,
respectively.  The latter value is already close the CNO equilibrium,
i.e., most of the initial \I{12}C and \I{16}O are converted into
\I{14}N.  A notable depletion of \I{12}C occurs only for ZAMS
rotational velocities of $\gtrsim200\,\kms$, but becomes as high as a
factor $\sim30$ for the fastest rotating models --- which is even
higher than the depletion factor of \I{12}C in the stellar center,
since the \I{12}C CNO equilibrium abundance is lower for lower
temperatures \cite[][]{AM93}.  The depletion of \I{16}O and the
enrichment of \I4{He} are correlated and become significant only for
stars with ZAMS rotational velocities of $\gtrsim300\,\kms$, the
stronger the higher the initial mass of the stars.  The coupling of
these two isotopes is due to the $\mu$-barrier: both, the destruction
of \I{16}O and the production of \I4{He} occur within or below the
$\mu$-barrier \cite[][]{HLW00:I}.

In the extremely rapidly spinning $20\,\Msun$ Model H20B the surface
helium enrichment becomes so large ($\gtrsim60\,\%$ by mass), at a
central hydrogen mass fraction of $\sim4\,\%$, that the star can be
considered a Wolf-Rayet (WR) star, with a correspondingly high mass
loss rate.  Even though the star experiences high mass loss during
that WR phase, it becomes a red supergiant after core hydrogen
exhaustion.  The high mass loss rate reduces the rotational velocity
and the luminosity of the star, as shown in
\FIGS{d3}{t-CNOvL} and \FIGFF{e3}{t-CNOvL}.

The strength of the surface enrichment is correlated with the time
when notable enrichment first appears during the evolution.  This is a
sign of an over-all increasing strength of the mixing with increasing
initial angular momentum.  It is interesting to note that the
$\mu$-turn in the HR diagram (\Sect{HburnHRD}) for the fast rotating
models also coincides to a ``kink'' in the evolution of the surface
abundances, especially for isotopes produced below the $\mu$-barrier,
namely \I{16}O and \I4{He}, and, to some extent, \I{14}N.  These
$\mu$-turns appear at an age of about $10\,\Myr$, $20\,\Myr$,
$15\,\Myr$, and $10\,\Myr$ for Models F12B, H12B, H15B, and H20B,
respectively (\Fig{t-CNOHe}).

\subsubsection{F, Na, Al, Mg}

Other interesting changes in the surface abundances due to
rotationally induced mixing are the depletion of fluorine (\I{19}F)
and enrichment in sodium (\I{23}{Na}), both the only stable isotopes
of these elements (\FIGS{b1-b3,c1-c3}{t-BFNaAl}).  An important
product of the magnesium-aluminum cycle \cite[][]{AM93} is the
production of the radioactive isotope \I{26}{Al} by
\I{25}{Mg}\Rpg\I{26}{Al} in the central convective core.  Typically,
it reaches a mass fraction of a few \Ep{-6} in the center for the
initial composition used in the present work.  This isotope has a half
life of $0.716\,\Myr$ and emits $\gamma$-rays due to its
$\beta$-decay, which has been observed in the interstellar medium of
our Galaxy \cite[][]{obe96}.  The surface abundance of \I{26}{Al} shows
the interplay of production, decay, and mixing
(\FIGS{d1-d3}{t-BFNaAl}): a steep initial increase in the abundance
appears when \I{26}{Al} is mixed to the surface within the first few
$\Myr$.  For the fast rotating models, the abundance slowly continues
to increase until the mixing between core and envelope is shut off at
the $\mu$-turn, and the further evolution of the \I{26}{Al} surface
abundance is mainly governed by its decay.  The drop of the \I{26}{Al}
surface abundance, as, e.g., seen in \FIG{d1}{t-BFNaAl}, corresponds
to an exponential decay.  The apparently different slopes in
{\FIGS{d1-d3}{t-BFNaAl}} are due to the different time axes.

The surface abundance of \I{26}{Al} strongly increases with the amount
of mixing, i.e., with the ZAMS rotational velocity of the star.  But
it is also a sensitive function of the initial mass: The fastest
rotating $20\,\Msun$ star (Model H20B) has an about two orders of
magnitude higher surface abundance than the fastest rotating
$12\,\Msun$ star (Model H12B).  This difference is even larger for the
slower rotators.

The larger surface enrichment in \I{26}{Al} in the more massive stars
is due mostly to their shorter lifetime.  More massive stars show a
comparable enrichment in products of central hydrogen burning as a
function of the \emph{fraction} of their lifetime, within a factor
two of less massive stars, as can be seen, e.g., from the surface
enrichment in helium (\FIG{d1-d3}{t-CNOHe}).  But during the shorter
\emph{absolute} lifetime, less \I{26}{Al} decays.

Several interesting changes in the surface isotopic ratios occur for
different elements as well.  They are shown in \cite{Heg98}.  In
contrast to the changes in the elemental abundances, these isotopic
ratios are not easily measurable in the atmospheres of hot stars, and
therefore cannot be used as observational tests.

\section{Post-central hydrogen burning}
\lSect{postH}

\subsection{Evolution in the HR diagram}
\lSect{HeHRD}

\pFig{HRD-DE}

\pFig{HRD-DEB}

\pFig{HRD-12B}

\pFig{HRD-15B}

\pFig{HRD-20B}

After core hydrogen exhaustion the stars discussed in the present work
become red supergiants.  They evolve to the the Hayashi line within a
few $10\,000\,\yr$.  In \Figs{HRD-DE} and \Figff{HRD-DEB}, the
evolutionary tracks in the HR diagram of the rotating models with a
ZAMS rotational velocity of $\sim200\,\kms$ and non-rotating models
are compared.  The tracks go from the ZAMS up to core collapse, except
for the lowest mass Models D10, D12, and E08 in \Fig{HRD-DE}, which
end before neon ignition.  The evolution of Model E25 is only shown
until it becomes a Wolf-Rayet star during central helium burning,
where it loses its hydrogen-rich envelope.  In \Figs{HRD-12B},
\Figff{HRD-15B}, and \Figff{HRD-20B} the evolution of stars with
different initial rotation rates are compared.

Since the rotating stars develop larger helium cores, they become more
luminous red supergiants than the non-rotating stars of equal mass.
This is shown in \Fig{HRD-DE} for models with inefficient
$\mu$-gradients, and in \Fig{HRD-12B}, for different initial
rotational velocities.  In the former case, the models also evolve to
notably lower effective temperatures than their non-rotating
counterparts.  These models have considerably larger helium cores and,
due to their stronger mass loss, a by far higher envelope to core mass
and mass to luminosity ratio.  Due to this, these stars, especially
the most luminous of them, become highly unstable to radial pulsations
\cite[][]{heg97}.

In models where the effect of $\mu$-gradients on rotationally induced
mixing is taken into account, the effective temperatures reached on
the Hayashi line are not significantly different from those of the
non-rotating models.  The increase in luminosity is much less
(\Figs{HRD-15B} and \Figff{HRD-20B}), except for the fastest rotators,
like Model H12B \cite[see][]{Heg98}.  This model does not become a red
supergiant immediately after core hydrogen exhaustion, but burns most
of the helium as a blue supergiant before it moves to the Hayashi
line.

Stars with initial masses of $\lesssim12\,\Msun$ show so-called ``blue
loops'', i.e., they evolve from red into blue supergiants
(\Figs{HRD-DE}, \Figff{HRD-DEB} and \Figff{HRD-12B}).  Stars with more
rotationally induced mixing evolve to lower effective temperatures and
to higher luminosities during their blue loops (\Figs{HRD-DE},
\Figff{HRD-DEB}, and \Figff{HRD-12B}).  The highly peculiar evolution
of the surface rotation rate and mass loss during a blue loop is
discussed in \cite[][]{HL98}.  Here we give in \Tab{rot} the average
radii rotational velocities and specific angular momenta during the
blue loop.  The maximum rotation rate, which usually is obtained
briefly after the transition to a blue supergiant, can be much higher
than that average (\Sect{RotVel}).

\subsection{First dredge-up abundances}

\pTab{1st}

\pFig{v-CNOHe-RSG}

Red supergiant have an extended surface convection zone which
comprises the major part of the hydrogen-rich envelope.  Since
convective mixing is much faster than rotationally induced mixing, the
convective envelope is essentially chemically homogeneous.  The
thermonuclear processed matter which becomes engulfed into the
convective envelope appears ``instantaneously'' at the surface,
compared to the evolutionary time-scale of the star.  This rapid
mixing of processed matter to the surface, which occurs also in
non-rotating stars, is known as a \emph{dredge-up}, and its first
occurrence after core hydrogen depletion is accordingly called the
\emph{first} dredge-up.

In \Fig{v-CNOHe-RSG} thin lines show, as a function of the ZAMS
rotational velocity, the surface mass fraction of \I{12}C, \I{14}N,
and \I{16}O relative to their initial values and that of \I4{He} at
core hydrogen exhaustion for three different initial masses for models
where rotationally induced mixing was assumed to be sensitive to
$\mu$-gradients.  Thick lines give the same quantities after the first
dredge-up.  The surface abundances of elements up to sodium after the
first dredge-up are also given in \Tab{1st}.

The slowly and non-rotating models show a significant enhancement in
\I{14}N due to the dredge-up.  In contrast, in the fast rotators the
changes of the surface abundances due to dredge-up are
small in comparison to the surface enrichment which already occurred
during central hydrogen burning.  That is, the envelope is already
well-mixed before the dredge-up.

Since, during the dredge-up, the convective envelope penetrates into
regions which were located below the $\mu$-barrier, the surface carbon
mass fraction is reduced by diluting the \I{12}C in the envelope with
the \I{12}C-depleted matter from deeper inside the star.  In a similar
way, \I{16}O is depleted and \I4{He} is enriched.  The bottom of the
convective envelope does not get hot enough for nuclear burning, i.e.,
the abundance changes at the surface during the dredge-up result only
from mixing of already processed matter into the convective envelope
as it extends to lower mass coordinates.

In the case of rotationally induced mixing, conversion of \I{12}C into
\I{14}N occurs above the $\mu$-barrier and therefore the composition
of matter originally located close to the surface of the stars can be
altered by first mixing it down, then processing it, and mixing it up
again.  Due to this, the surface enrichment in the faster rotating
stars is stronger, both before and after the dredge-up, than in the
slowly or non-rotating stars.  The stronger depletion of \I{16}O and
enrichment of \I4{He} for the faster rotating stars is mainly because
in those the rotationally induced mixing can overcome the
$\mu$-barrier for a longer time during core hydrogen burning than in
the slow rotators.

\subsection{Second dredge-up abundances}
\lSect{2nd-dredge}

\pTab{2nd}

After core helium exhaustion, the bottom of the convective envelope of
the red supergiants moves to lower mass coordinates.  Thus, it can
transport matter which has been processed by the hydrogen shell to the
surface of the star.  However, in most of the models considered here,
the convective envelope does not penetrate much deeper during the
second dredge-up than it did during the first dredge-up.
Additionally, not much mixing occurs in the radiative part of the
envelope during core helium burning, and the temperatures are mostly
too low for significant thermonuclear processing.  Thus, with the
exception of the lowest masses (e.g., Models D10 and E08), no strong
changes in the surface abundances occur during the second dredge-up.
In stars which experience a blue loop, the processing can be slightly
stronger, since during this loop the envelope contracts and a larger
part of it can become sufficiently hot for thermonuclear reactions to
occur.

The surface abundances of the most important isotopes after the second
dredge-up are given in \Tab{2nd}.  Since in the case of the second
dredge-up the mixing in the convective envelope is also fast enough to
make it chemically homogeneous, \Tab{2nd} gives the presupernova
abundances of the entire envelope.

\subsection{Rotational velocities}
\lSect{RotVel}

\pFig{t-vTeff-E12B}

\pTab{rot}

The majority of the stellar models investigated here expand to red
supergiants after core hydrogen depletion.  The expansion of the
stellar envelope increases the moment of inertia and the angular
velocity of the rigidly rotating convective envelope of the star
decreases considerably.  This is the most important effect
contributing to the slow-down of the star's rotation when evolving
from the main sequence to the red supergiant regime.  Additionally,
strong mass loss sets in during the red supergiant phase, and the
corresponding loss of angular momentum due to stellar winds becomes
important.  This is also true for the models of lower initial mass, in
which angular momentum loss was almost negligible during central
hydrogen burning.  Since the rise of the moment of inertia is much
larger than the growth in radius \cite[][]{HL98}, the surface
rotational velocity decreases considerably and drops below $1\,\kms$
in most models (\Tab{rot}, {\Fig{t-vTeff-E12B}}).  During the second
dredge-up, i.e., after core helium exhaustion, the models become even
more luminous and extended (\Tab{rot}).

Models with an initial mass of $12\,\Msun$ or below, experiencing a
blue loop, show a remarkable spin-up during this phase and can lose
the major part of their total angular momentum \cite[see][]{HL98}.
This explains the small difference in the angular velocity of the
$12\,\Msun$ models compared to the corresponding $15\,\Msun$ models,
which do not develop blue loops, in relation to the differences
between other models with same input physics but different initial
mass (see e.g., Models E12B, E15B, and E20B in \Tab{rot}).

Blue supergiants born out of such a transition can rotate close to
their critical rotation rate.  This fast rotation occurs because the
envelope angular momentum is concentrated in a small mass fraction at
the surface of the star.  Due to mass loss the rapidly rotating layers
get lost soon and the star slows down quickly \cite[see][]{HL98}.

Model H12B evolves as a blue supergiant during the first part of
 core helium burning.  Since it has a large initial angular
momentum and does not experience any significant previous angular
momentum loss, it evolves close to critical rotation during this
phase.  Even the enhancement of the mass loss rate due to rotation
does not cause a sufficiently high angular momentum loss to slow it
down from critical rotation.

Models E25 and H20B become Wolf-Rayet stars during central helium
burning.  The evolution of Model E25 has been followed until core
collapse.  The strong mass loss during the Wolf-Rayet phase rapidly
slows down the rotation of the helium envelope.  None of the
instabilities considered in the present work is fast enough to keep it
in rigid rotation with the core or the rest of the helium-rich layers.
Thus strong differential rotation occurs in the envelope and the core
is not slowed down significantly.  The slower rotation of the core
compared to other models \cite[see][]{HLW00:I} is mostly due to the loss of
angular momentum during core hydrogen burning.

\section{Observational evidence}
\lSect{obs}


There is a considerable body of observations of abundance anomalies in
massive stars. In particular, helium and CN-abundances of main
sequence~O and B~stars by \cite{GL92,Her94,vra00} have been used to
calibrate the mixing efficiencies for our models \cite[][]{HLW00:I},
which thus reproduced this \emph{on average}.  For a moderate initial
rotation rate, our models obtain a rather insignificant helium
enrichment ($\sim$1\% by mass), a carbon depletion of less than
$50\,\%$, and a nitrogen enhancement of more than a factor two
(\Tab{TAMS}).  Although the observational facts about OB~main sequence
stars are not completely unambigous, the quoted numbers appear to be
supported by recent empirical studies
\cite[][]{MLD98,SH98,her99,kor99,MLD99}.  However, our main sequence
models provide the following {\em testable predictions}:
\begin{enumerate}
\item
\emph{Stronger surface anomalies should occur for larger rotation
velocities, \vrot.}\/ A test of this feature is complicated by two
things, namely that not \vrot but only $\vrot \sin i$ is accessible to
observations, where $i$ is the inclination angle, and that abundance
determinations are often only possible for stars with relatively small
values of $\vrot \sin i$ \cite[cf.][]{GL92}.  Nevertheless, some tentative
support for a positive correlation of enrichment and rotation comes
from \cite{MLD99} and \cite{her99}, although other interpretations of
the observational data appear possible as well.

\item
\emph{Stronger surface anomalies should occur for more luminous
stars.}\/  This seems in fact to be the case, according to \cite{MLD99}
and \cite{GL92}.

\item
\emph{Lithium, beryllium and boron are considerably depleted.}\/ While
there is no observational data on the first two elements for hot
luminous stars, the boron abundance in B~star has been investigated by
\cite{VLL96}, and by \cite{cun97}.  \cite{FLV96} interpreted the
boron depletion in B~stars which at the same time are not
enriched in nitrogen as evidence for rotational mixing.
\end{enumerate}

There is also evidence from more evolved luminous stars supporting the
scenario of rotational mixing of the kind obtained by our models.
However, an additional problem appears in this case, namely that ---
for stars in the mass range considered here --- the post main sequence
radius or temperature evolution can not yet be reliably predicted by
stellar evolution theory \cite[][]{LM95}.  Therefore, it is often not
unambiguous which fraction of an abundance anomaly in an evolved
massive star might be attributed to a possible previous convective
dredge-up which would have occurred were the star a red supergiant in
the past.

However, \cite{Ven95:1,Ven95:2,Ven99} has analyzed abundances in
Galactic and Magellanic Cloud A~supergiants and found results which
are incompatible with standard evolution with or without red
supergiant dredge-up but rather call for an additional enrichment
process like rotationally induced mixing.

Further abundance anomalies which might be related to rotational
mixing are the strong signature of CNO processing in the progenitor of
SN~1987A \cite[][]{fra89}, small \I{12}C/\I{13}C-ratios in red giants
\cite[][]{bri95}, sodium enrichment in cool supergiants
\cite[][]{boy88,Hil97,HBS97}, and simultaneous carbon and nitrogen
enrichment in several Wolf-Rayet stars \cite[][for more
details]{CSW95,Lan91C}.

While perhaps none of the quoted observational evidences is completely
convincing by itself, their huge amount gives strong support for the
idea that a non-standard (i.e., non-convective) enrichment process
must in fact operate in massive stars.  As the rotational mixing
during core hydrogen burning can also increase the luminosity-to-mass
ratio, the so called mass discrepancy, i.e., the finding that masses
of massive main sequence stars derived through spectroscopy or by
constraints from radiation driven wind models are on average smaller
than masses derived from stellar evolutionary tracks
\cite[cf.][]{her92,HVM98}, gives further support to extra mixing in
massive stars.

The only scenario presently competing with the rotational mixing
scenario is that of close binary mass transfer.  In fact, many massive
stars are members of close binary systems, and one should expect mass
transfer to contribute to the observed abundance anomalies in luminous
stars.  In a recent comprehensive study of massive close binary
evolutionary models, \cite{WL99} and \cite{WLB00} find in fact that
the enrichment of the accreting star in a binary may closely resemble
the enrichment produced by rotational mixing in the models presented
here.  In particular, the CNO pattern, and the helium and sodium
enhancements are found to be very similar.  Moreover, those binary
components which accrete significant amounts of mass from their
companion end up in rather wide orbits, with companions of much lower
mass.  That is, after the supernova explosions of the companions, the
accretion stars may not be easily recognizable by a large space
velocity.  Due to the accretion of angular momentum they might also be
rapid rotators, which may lead to a positive correlation of enrichment
and rotational velocity.

Nevertheless, we identify two distinguishing characteristics of the
rotation and the binary enrichment scenarios. First, as already
anticipated by \cite{FLV96}, boron depleted main sequence stars which
are \emph{not} nitrogen enriched are only produced by rotation.  While
\cite{FLV96} used estimates to exclude such abundance pattern for the
binary scenario, this is now confirmed in detail by the recent grid of
binary models of \cite{WLB00}.  Second, the rotational mixing scenario
predicts a stronger enrichment for more luminous stars (see above).
In contrast, the binary models of \cite{WL99} and \cite{WLB00}, which
cover a luminosity range of $\log L/\Lsun \simeq 4.5\ldots6.0$, do not
show this trend.

As there is empirical evidence for both distinguishing predictions of
the rotational mixing scenario \cite[cf.][]{FLV96,MLD99}, the contribution of
the binary scenario seems not to be the dominant one.  However, more
observations are clearly needed to prove its significance.  We suggest
that in particular the combined study of boron and CNO elements in hot
stars will allow to draw more stringent conclusions, and that a
simultaneous derivation of the rotational and also of the radial
velocities would be very valuable to provide further clues as to the
relative importance of the mass transfer and the rotational mixing
scenario for an understanding of the multitude of abundance
peculiarities in massive stars.

\section{Conclusions}
\lSect{conclude}

Observations show that evolved massive main-sequence stars have
nuclear processed matter at their surfaces (\Sect{obs}).  Our models
show that these surface abundance anomalies can be well explained by
rotationally induced mixing.  While the mixing processes we consider
are gauged to reproduce the \emph{average} surface enrichment of
nitrogen and helium, they also reproduce distinctive abundance trends
that are not expected in competing binary scenarios, where the surface
anomalies are due to mass transfer from the companion.  Specifically,
our models produce a boron depletion without nitrogen enrichment and
an increasing nitrogen abundance with increasing luminosity.

Due to the strong temperature dependence of the destruction reactions,
light isotopes will essentially be either preserved or completely
destroyed at the surface of non-rotating stars, depending on the
strength of the mass loss.  Rotationally induced mixing, on the other
hand, leads to a gradual depletion.  \emph{Rotationally reduced
depletion} can also preserve some of the light elements in cases where
a non-rotating stars would show complete depletion, although the
resulting depletion might still be quite considerable (\Tab{TAMS}).

Due to rotationally induced mixing, the chemical signature of hydrogen
burning is apparent at the stellar surface already on the main
sequence.  Prominent features are enrichments of \I4{He}, \I{13}C,
\I{14}N, \I{17}O, \I{23}{Na}, and \I{26}{Al}, while \I{12}C, \I{15}N,
\I{16}O, \I{18}O, and \I{19}F are depleted (\Tab{TAMS}).  The faster
the initial rotation the stronger the resulting mixing and the longer
the mixing between core and envelope can be maintained until a
sufficiently large $\mu$-barrier forms which suppresses further mixing
(\Fig{t-BFNaAl}).

The mixing between core and envelope increases the mean molecular
weight of the star and leads to larger convective core and higher
luminosity at the end of central hydrogen burning, i.e., the
main-sequence band is broadened (\Figs{HRD-DE} and \Figff{HRD-DEB}).
The larger fuel supply, but even more so the lower initial luminosity
of the rotating stars, helps to prolong the time they spend on the
main sequence (\Fig{t-CNOvL}).  As the track in the HR diagram can be
significantly changed by rotation, a given position on the main
sequence band does not unambiguously correspond to one initial mass
(\Sect{HburnHRD}), i.e., there is no unique mass-luminosity relation
for main sequence stars.  Despite mass loss and expansion, the stellar
rotation velocity stays about constant on the main sequence
(\Fig{t-CNOvL}).

The \emph{first} dredge-up changes the surface abundances by a factor
which is smaller the larger the initial rotation rate is, since the
fast rotators are already well mixed at the end of central hydrogen
burning (\Fig{v-CNOHe-RSG}).  Even after the dredge-up, the fast
rotators maintain chemical peculiarities, as, e.g., a significantly
stronger CNO processing.  The \emph{second} dredge-up does hardly
affect the surface abundances, except in the two low-mass cases where
the surface convection zone penetrates into the helium shell.  Some
small abundance changes are also present in the models that undergo a
blue loop, since there the hydrogen shell burning is more extended in
mass.  However, after the blue loop, these red supergiants are
rotating slower by a factor $\sim$2--3, compared to red supergiants
without a previous blue loop.

\acknowledgments

We are grateful to Danny Lennon, Kim Venn, and Stan Woosley for many
useful discussions.  This work was supported by the National Science
Foundation (AST 97-31569, INT-9726315), the Deutsche
Forschungsgemeinschaft (La~587/15, 16) and the Alexander von
Humboldt-Stiftung (FLF-1065004).  AH was, in part, supported by a
``Doktorandenstipendium aus Mitteln des 2. Hochschulprogramms''.

\bibliographystyle{mybibstyle}


\clearpage
\onecolumn


\begin{landscape}
\begin{table}
\centering
{\small
\begin{tabular}[t]{lRRRRRRRRRRRRRRR} 
\hline\hline
model
& \multicolumn{1}{r}{X(\I4{He})}
& \multicolumn{1}{r}{[\I3{He}]}
& \multicolumn{1}{r}{[\El{Li}]}
& \multicolumn{1}{r}{[\El{Be}]}
& \multicolumn{1}{r}{[\El{B}]}
& \multicolumn{1}{r}{[\I{12}C]}
& \multicolumn{1}{r}{[\I{13}C]}
& \multicolumn{1}{r}{[\I{14}N]}
& \multicolumn{1}{r}{[\I{15}N]}
& \multicolumn{1}{r}{[\I{16}O]}
& \multicolumn{1}{r}{[\I{17}O]}
& \multicolumn{1}{r}{[\I{18}O]}
& \multicolumn{1}{r}{[\I{19}F]}
& \multicolumn{1}{r}{[\I{23}{Na}]}
\\
\hline
D10  &  0.280 &  0.000 &  0.000 &  0.000 &  0.000 &  0.000 &  0.000 &  0.000 &  0.000 &  0.000 &  0.000 &  0.000 &  0.000 &  0.000  \\
D12  &  0.280 &  0.000 &  0.000 &  0.000 &  0.000 &  0.000 &  0.000 &  0.000 &  0.000 &  0.000 &  0.000 &  0.000 &  0.000 &  0.000  \\
D15  &  0.280 &  0.000 &  0.000 &  0.000 &  0.000 &  0.000 &  0.000 &  0.000 &  0.000 &  0.000 &  0.000 &  0.000 &  0.000 &  0.000  \\
D20  &  0.280 &  0.000 & -6.006 & -0.916 & -0.001 &  0.000 &  0.000 &  0.000 &  0.000 &  0.000 &  0.000 &  0.000 &  0.000 &  0.000  \\
D25  &  0.280 &  0.000 & -0.006 & -0.006 & -0.001 &  0.000 &  0.000 &  0.000 &  0.000 &  0.000 &  0.000 &  0.000 &  0.000 &  0.000  \\
\hline
E08  &  0.280 &  0.010 & -0.456 & -0.456 & -0.418 & -0.016 &  0.252 &  0.053 & -0.049 & -0.001 &  0.377 & -0.038 & -0.002 &  0.007  \\
E10  &  0.281 & -0.018 & -0.633 & -0.633 & -0.558 & -0.033 &  0.358 &  0.110 & -0.085 & -0.003 &  0.634 & -0.071 & -0.006 &  0.020  \\
E12  &  0.282 & -0.042 & -0.755 & -0.754 & -0.637 & -0.044 &  0.413 &  0.149 & -0.108 & -0.006 &  0.735 & -0.095 & -0.011 &  0.031  \\
E15  &  0.286 & -0.084 & -1.053 & -1.052 & -0.833 & -0.069 &  0.510 &  0.238 & -0.164 & -0.013 &  0.941 & -0.150 & -0.023 &  0.064  \\
E20  &  0.293 & -0.125 & -1.457 & -1.451 & -1.049 & -0.090 &  0.577 &  0.323 & -0.220 & -0.027 &  1.069 & -0.207 & -0.043 &  0.102  \\
E25  &  0.303 & -0.149 & -5.555 & -3.180 & -1.237 & -0.105 &  0.566 &  0.408 & -0.246 & -0.047 &  1.196 & -0.236 & -0.073 &  0.151  \\
\hline
G12  &  0.280 &  0.001 & -1.569 & -0.779 & -0.184 &  0.000 &  0.005 &  0.000 & -0.001 &  0.000 &  0.000 & -0.001 &  0.000 &  0.000  \\
\hline
F12  &  0.323 & -0.232 & -8.662 & -4.719 & -1.600 & -0.251 &  0.612 &  0.587 & -0.409 & -0.076 &  1.571 & -0.423 & -0.125 &  0.265  \\
\hline
G12B &  0.280 & -0.013 & -3.002 & -1.532 & -0.491 & -0.012 &  0.231 &  0.036 & -0.043 &  0.000 &  0.147 & -0.035 & -0.001 &  0.001  \\
G15B &  0.280 & -0.027 & -3.177 & -1.685 & -0.562 & -0.018 &  0.307 &  0.052 & -0.063 &  0.000 &  0.239 & -0.053 & -0.002 &  0.002  \\
G20B &  0.280 & -0.044 & -3.496 & -1.910 & -0.671 & -0.024 &  0.393 &  0.064 & -0.090 &  0.000 &  0.307 & -0.077 & -0.003 &  0.002  \\
\hline
E12B &  0.284 & -0.201 & -8.017 & -4.271 & -1.503 & -0.212 &  0.651 &  0.413 & -0.404 & -0.014 &  1.426 & -0.394 & -0.051 &  0.094  \\
E15B &  0.284 & -0.264 & -8.423 & -4.657 & -1.680 & -0.237 &  0.722 &  0.438 & -0.486 & -0.017 &  1.464 & -0.482 & -0.062 &  0.097  \\
E20B &  0.289 & -0.311 & -8.734 & -5.090 & -1.908 & -0.239 &  0.785 &  0.470 & -0.558 & -0.031 &  1.519 & -0.563 & -0.089 &  0.124  \\
\hline
F12B &  0.318 & -0.466 & -9.983 & -10.16 & -3.596 & -0.572 &  0.714 &  0.730 & -0.970 & -0.115 &  2.048 & -1.253 & -0.305 &  0.396  \\
F15B &  0.331 & -0.624 & -9.992 & -10.45 & -3.810 & -0.714 &  0.562 &  0.783 & -0.986 & -0.135 &  1.971 & -1.383 & -0.324 &  0.397  \\
F20B &  0.338 & -0.671 & -9.890 & -10.92 & -4.140 & -0.645 &  0.651 &  0.793 & -1.002 & -0.155 &  1.926 & -1.448 & -0.361 &  0.392  \\
\hline
H12B &  0.461 & -0.984 & -9.743 & -15.64 & -8.257 & -1.539 & -0.075 &  1.000 & -0.946 & -0.455 &  2.088 & -3.397 & -1.229 &  0.635  \\
H15B &  0.521 & -1.259 & -9.778 & -14.41 & -9.931 & -1.532 & -0.062 &  1.040 & -0.910 & -0.600 &  1.906 & -3.933 & -1.579 &  0.654  \\
H20B &  0.654 & -1.876 & -9.915 & -13.02 & -16.05 & -1.460 &  0.010 &  1.074 & -0.887 & -0.808 &  1.628 & -4.414 & -1.904 &  0.668  \\
\hline
\end{tabular}
}
\caption{
Surface abundances at core hydrogen exhaustion for the indicated model
sequences.  Given are the {\I4{He}} mass fraction, and the 
logarithm of the ratios of the {\I3{He}}, lithium, boron, beryllium,
{\I{12}C}, {\I{13}C}, {\I{14}N}, {\I{15}N}, {\I{16}O}, {\I{17}O},
{\I{18}O}, {\I{19}F}, and {\I{23}{Na}} mass fractions relative to
their initial values.
\lTab{TAMS}}
\end{table}
\end{landscape}

\clearpage

\begin{landscape}
\begin{table}
\centering
{\small
\begin{tabular}[t]{lRRRRRRRRRRRRRRR} 
\hline\hline
model
& \multicolumn{1}{r}{X(\I4{He})}
& \multicolumn{1}{r}{[\I3{He}]}
& \multicolumn{1}{r}{[\El{Li}]}
& \multicolumn{1}{r}{[\El{Be}]}
& \multicolumn{1}{r}{[\El{B}]}
& \multicolumn{1}{r}{[\I{12}C]}
& \multicolumn{1}{r}{[\I{13}C]}
& \multicolumn{1}{r}{[\I{14}N]}
& \multicolumn{1}{r}{[\I{15}N]}
& \multicolumn{1}{r}{[\I{16}O]}
& \multicolumn{1}{r}{[\I{17}O]}
& \multicolumn{1}{r}{[\I{18}O]}
& \multicolumn{1}{r}{[\I{19}F]}
& \multicolumn{1}{r}{[\I{23}{Na}]}
\\
\hline
D10  &  0.301 & -0.113 & -1.133 & -1.133 & -1.087 & -0.212 &  0.490 &  0.492 & -0.318 & -0.038 &  1.346 & -0.304 & -0.060 &  0.170  \\
D12  &  0.305 & -0.160 & -1.043 & -1.043 & -1.019 & -0.216 &  0.497 &  0.511 & -0.324 & -0.045 &  1.298 & -0.314 & -0.067 &  0.183  \\
D15  &  0.312 & -0.203 & -1.147 & -1.147 & -1.091 & -0.219 &  0.509 &  0.538 & -0.332 & -0.056 &  1.293 & -0.326 & -0.083 &  0.204  \\
D20  &  0.334 & -0.250 & -7.935 & -3.416 & -1.309 & -0.226 &  0.527 &  0.589 & -0.346 & -0.079 &  1.229 & -0.348 & -0.108 &  0.241  \\
D25  &  0.354 & -0.273 & -1.715 & -1.714 & -1.435 & -0.228 &  0.542 &  0.631 & -0.353 & -0.104 &  1.226 & -0.361 & -0.138 &  0.275  \\
\hline
E08  &  0.363 & -0.183 & -1.326 & -1.326 & -1.251 & -0.336 &  0.546 &  0.669 & -0.475 & -0.100 &  1.576 & -0.498 & -0.140 &  0.305  \\
E10  &  0.388 & -0.300 & -1.572 & -1.571 & -1.409 & -0.375 &  0.544 &  0.715 & -0.515 & -0.123 &  1.502 & -0.561 & -0.168 &  0.339  \\
E12  &  0.403 & -0.370 & -1.758 & -1.757 & -1.520 & -0.394 &  0.558 &  0.741 & -0.544 & -0.140 &  1.446 & -0.606 & -0.187 &  0.360  \\
E15  &  0.423 & -0.436 & -2.030 & -2.025 & -1.697 & -0.409 &  0.582 &  0.773 & -0.577 & -0.167 &  1.409 & -0.659 & -0.219 &  0.388  \\
E20  &  0.447 & -0.492 & -2.507 & -2.484 & -1.948 & -0.413 &  0.613 &  0.810 & -0.611 & -0.207 &  1.365 & -0.715 & -0.267 &  0.419  \\
E25  &  0.627 & -0.945 & -6.780 & -4.776 & -2.565 & -0.751 &  0.382 &  0.981 & -0.794 & -0.447 &  1.463 & -1.156 & -0.732 &  0.562  \\
\hline
G12B &  0.310 & -0.264 & -4.764 & -3.017 & -1.579 & -0.301 &  0.630 &  0.574 & -0.509 & -0.053 &  1.417 & -0.515 & -0.087 &  0.208  \\
G15B &  0.317 & -0.309 & -4.529 & -2.873 & -1.507 & -0.300 &  0.645 &  0.592 & -0.521 & -0.063 &  1.370 & -0.535 & -0.100 &  0.221  \\
G20B &  0.337 & -0.350 & -4.874 & -3.132 & -1.646 & -0.288 &  0.688 &  0.628 & -0.539 & -0.087 &  1.332 & -0.560 & -0.127 &  0.256  \\
\hline
E12B &  0.314 & -0.410 & -9.111 & -5.631 & -2.434 & -0.475 &  0.647 &  0.656 & -0.765 & -0.064 &  1.624 & -0.850 & -0.127 &  0.246  \\
E15B &  0.322 & -0.468 & -8.756 & -5.734 & -2.479 & -0.461 &  0.701 &  0.672 & -0.800 & -0.078 &  1.606 & -0.909 & -0.147 &  0.264  \\
E20B &  0.345 & -0.514 & -7.905 & -6.090 & -2.736 & -0.432 &  0.743 &  0.702 & -0.819 & -0.106 &  1.578 & -0.950 & -0.185 &  0.295  \\
\hline
F12B &  0.358 & -0.571 & -9.538 & -11.42 & -4.401 & -0.711 &  0.626 &  0.818 & -1.023 & -0.177 &  2.037 & -1.577 & -0.395 &  0.464  \\
F15B &  0.369 & -0.747 & -8.786 & -8.246 & -4.581 & -0.866 &  0.482 &  0.841 & -1.038 & -0.178 &  1.945 & -1.757 & -0.379 &  0.443  \\
F20B &  0.394 & -0.810 & -7.482 & -6.096 & -4.939 & -0.785 &  0.559 &  0.863 & -1.024 & -0.215 &  1.880 & -1.769 & -0.434 &  0.452  \\
\hline
H12B &  0.481 & -1.072 & -9.421 & -12.97 & -9.033 & -1.550 & -0.085 &  1.007 & -0.945 & -0.478 &  2.065 & -3.665 & -1.264 &  0.640  \\
H20B &  0.700 & -2.080 & -6.555 & -4.861 & -18.00 & -1.447 &  0.023 &  1.080 & -0.883 & -0.860 &  1.570 & -4.549 & -1.982 &  0.674  \\
\hline
\end{tabular}
}
\caption{
Surface abundances after the first dredge-up for the indicated model
sequences.  Given are the {\I4{He}} mass fraction, and the 
logarithm of the ratios of the {\I3{He}}, lithium, boron, beryllium,
{\I{12}C}, {\I{13}C}, {\I{14}N}, {\I{15}N}, {\I{16}O}, {\I{17}O},
{\I{18}O}, {\I{19}F}, and {\I{23}{Na}} mass fractions relative to
their initial values.
\lTab{1st}}
\end{table}
\end{landscape}

\clearpage

\begin{table}
\centering
\begin{tabular}[t]{lRRRRRRRRR} 
\hline\hline
model
& \multicolumn{1}{r}{X(\El{He})}
& \multicolumn{1}{r}{[\El{Li}]}
& \multicolumn{1}{r}{[\El{Be}]}
& \multicolumn{1}{r}{[\El{B}]}
& \multicolumn{1}{r}{[\El{C}]}
& \multicolumn{1}{r}{[\El{N}]}
& \multicolumn{1}{r}{[\El{O}]}
& \multicolumn{1}{r}{[\El{F}]}
& \multicolumn{1}{r}{[\El{Na}]}
\\
\hline
D10  &  0.382 &  -1.21 &  -1.21 &  -1.17 &  -0.25 &   0.64 &  -0.09 &  -0.12 &   0.28  \\
D12  &  0.306 &  -1.26 &  -1.26 &  -1.19 &  -0.21 &   0.52 &  -0.04 &  -0.07 &   0.19  \\
D15  &  0.312 &  -1.17 &  -1.17 &  -1.11 &  -0.20 &   0.54 &  -0.05 &  -0.08 &   0.20  \\
D20  &  0.334 &  -7.92 &  -3.43 &  -1.32 &  -0.21 &   0.59 &  -0.08 &  -0.11 &   0.24  \\
D25  &  0.354 &  -1.73 &  -1.73 &  -1.44 &  -0.21 &   0.63 &  -0.10 &  -0.14 &   0.28  \\
\hline
E08  &  0.395 &  -1.49 &  -1.49 &  -1.39 &  -0.34 &   0.71 &  -0.12 &  -0.17 &   0.34  \\
E10  &  0.388 &  -1.71 &  -1.71 &  -1.54 &  -0.35 &   0.72 &  -0.12 &  -0.17 &   0.34  \\
E12  &  0.403 &  -1.87 &  -1.87 &  -1.63 &  -0.37 &   0.74 &  -0.13 &  -0.19 &   0.36  \\
E15  &  0.423 &  -2.08 &  -2.07 &  -1.74 &  -0.37 &   0.77 &  -0.16 &  -0.22 &   0.39  \\
E20  &  0.450 &  -2.60 &  -2.55 &  -2.03 &  -0.39 &   0.82 &  -0.20 &  -0.27 &   0.42  \\
\hline
G12B &  0.312 &  -4.78 &  -3.04 &  -1.60 &  -0.28 &   0.58 &  -0.05 &  -0.09 &   0.21  \\
G15B &  0.317 &  -4.56 &  -2.90 &  -1.53 &  -0.27 &   0.59 &  -0.06 &  -0.10 &   0.22  \\
G20B &  0.337 &  -4.90 &  -3.16 &  -1.66 &  -0.25 &   0.63 &  -0.08 &  -0.13 &   0.26  \\
\hline
E12B &  0.315 &  -9.13 &  -5.65 &  -2.45 &  -0.43 &   0.66 &  -0.06 &  -0.13 &   0.25  \\
E15B &  0.323 &  -8.77 &  -5.76 &  -2.50 &  -0.41 &   0.67 &  -0.07 &  -0.15 &   0.27  \\
E20B &  0.346 &  -7.91 &  -6.11 &  -2.75 &  -0.37 &   0.70 &  -0.10 &  -0.19 &   0.30  \\
\hline
F12B &  0.358 &  -9.43 & -11.47 &  -4.45 &  -0.64 &   0.82 &  -0.15 &  -0.40 &   0.46  \\
F15B &  0.369 &  -8.80 &  -8.32 &  -4.61 &  -0.78 &   0.84 &  -0.16 &  -0.38 &   0.44  \\
F20B &  0.394 &  -7.49 &  -6.16 &  -4.96 &  -0.70 &   0.86 &  -0.20 &  -0.43 &   0.45  \\
\hline
H12B &  0.490 &  -9.42 & -11.55 &  -9.08 &  -1.42 &   1.01 &  -0.43 &  -1.28 &   0.64  \\
\hline
\end{tabular}
\caption{
Surface abundances after the second dredge-up for the indicated model
sequences.  Given are the helium mass fraction, and the
logarithm of the ratios of the lithium, boron, beryllium, nitrogen,
carbon, oxygen, fluorine, and sodium mass fractions relative to their
initial values.
\lTab{2nd}}
\end{table}

\clearpage

\begin{landscape}
\begin{table}
\centering
\begin{tabular}[t]{l|RRRR|RRRR|RRRRRR|RRR} 
\hline\hline
& \multicolumn{4}{c|}{ZAMS}
& \multicolumn{4}{c|}{RSG}
& \multicolumn{6}{c|}{blue loop}
& \multicolumn{3}{c}{presupernova}
\\
\hline
model
& \multicolumn{1}{C}{R}
& \multicolumn{1}{C}{v}
& \multicolumn{1}{C}{\jeq}
& \multicolumn{1}{C|}{\tMS}
& \multicolumn{1}{C}{\Av{R}}
& \multicolumn{1}{C}{\Av{v}}
& \multicolumn{1}{C}{\Av{\jeq}}
& \multicolumn{1}{C|}{\tRSG}
& \multicolumn{1}{C}{\Av{R}}
& \multicolumn{1}{C}{\Av{v}}
& \multicolumn{1}{C}{\vmax}
& \multicolumn{1}{C}{\Av{\jeq}}
& \multicolumn{1}{C}{\jeqm}
& \multicolumn{1}{C|}{\tBSG}
& \multicolumn{1}{C}{R}
& \multicolumn{1}{C}{v}
& \multicolumn{1}{C}{\jeq}
\\
(mass)
& \multicolumn{1}{C}{\Rsun}
& \multicolumn{1}{C}{\kms}
& \multicolumn{1}{C}{}
& \multicolumn{1}{C|}{\Myr}
& \multicolumn{1}{C}{\Rsun}
& \multicolumn{1}{C}{\kms}
& \multicolumn{1}{C}{}
& \multicolumn{1}{C|}{\Myr}
& \multicolumn{1}{C}{\Rsun}
& \multicolumn{1}{C}{\kms}
& \multicolumn{1}{C}{\kms}
& \multicolumn{1}{C}{}
& \multicolumn{1}{C}{}
& \multicolumn{1}{C|}{\Myr}
& \multicolumn{1}{C}{\Rsun}
& \multicolumn{1}{C}{\kms}
& \multicolumn{1}{C}{}
\\
\hline
D10  &   3.90 &       0 &   0.00 &  18.72 &     254 &   0.00 &   0.00 &   2.27 &      39 &       0 &       0 &       0 &       0 &   0.96 &     735 &   0.00 &   0.00  \\
D12  &   4.32 &       0 &   0.00 &  13.85 &     344 &   0.00 &   0.00 &   1.24 &      42 &       0 &       0 &       0 &       0 &   0.81 &     486 &   0.00 &   0.00  \\
D15  &   4.89 &       0 &   0.00 &  10.10 &     520 &   0.00 &   0.00 &   1.21 &       - &       - &       - &       - &       - &      - &     711 &   0.00 &   0.00  \\
D20  &   5.74 &       0 &   0.00 &   6.97 &     842 &   0.00 &   0.00 &   0.74 &       - &       - &       - &       - &       - &      - &  1\,043 &   0.00 &   0.00  \\
D25  &   6.49 &       0 &   0.00 &   5.90 &  1\,207 &   0.00 &   0.00 &   0.42 &       - &       - &       - &       - &       - &      - &  1\,393 &   0.00 &   0.00  \\
\hline
E08  &   3.54 &     205 &   5.08 &  36.97 &     202 &   2.58 &   3.08 &   3.64 &      51 &      57 &      74 &      19 &      25 &   2.06 &     704 &   0.19 &   0.96  \\
E10  &   4.00 &     208 &   5.84 &  25.25 &     338 &   1.38 &   2.89 &   1.90 &      61 &      53 &     111 &      20 &      65 &   1.13 &     700 &   0.16 &   0.78  \\
E12  &   4.42 &     208 &   6.44 &  18.78 &     486 &   0.83 &   2.58 &   1.58 &      92 &      31 &     126 &      17 &      98 &   0.51 &     902 &   0.06 &   0.39  \\
E15  &   5.00 &     208 &   7.29 &  13.70 &     762 &   0.32 &   1.66 &   1.26 &       - &       - &       - &       - &       - &      - &  1\,157 &   0.05 &   0.37  \\
E20  &   5.84 &     202 &   8.26 &   9.60 &  1\,208 &   0.11 &   0.87 &   0.67 &       - &       - &       - &       - &       - &      - &  1\,608 &   0.01 &   0.11  \\
\hline
G12B &   4.35 &      99 &   3.03 &  13.63 &     361 &   0.77 &   1.76 &   1.42 &      44 &      38 &      61 &      11 &      29 &   0.63 &     489 &   0.29 &   1.01  \\
G15B &   4.93 &     102 &   3.52 &   9.91 &     520 &   0.48 &   1.68 &   1.27 &       - &       - &       - &       - &       - &      - &     739 &   0.20 &   1.02  \\
G20B &   5.78 &     103 &   4.19 &   7.04 &     826 &   0.23 &   1.33 &   0.83 &       - &       - &       - &       - &       - &      - &  1\,061 &   0.09 &   0.63  \\
\hline
E12B &   4.45 &     206 &   6.40 &  14.30 &     367 &   1.56 &   3.54 &   1.45 &      51 &      65 &     124 &      22 &     106 &   0.62 &     542 &   0.43 &   1.64  \\
E15B &   5.03 &     206 &   7.26 &  10.49 &     537 &   0.88 &   3.18 &   1.30 &       - &       - &       - &       - &       - &      - &     773 &   0.34 &   1.84  \\
E20B &   5.88 &     201 &   8.29 &   7.50 &     906 &   0.36 &   2.22 &   0.81 &       - &       - &       - &       - &       - &      - &  1\,110 &   0.13 &   1.01  \\
\hline
F12B &   4.64 &     328 &  10.63 &  18.68 &     413 &   2.02 &   5.11 &   1.41 &      66 &      52 &     120 &      22 &     130 &   0.55 &     660 &   0.25 &   1.13  \\
F15B &   5.22 &     323 &  11.76 &  13.02 &     574 &   1.11 &   4.35 &   1.26 &       - &       - &       - &       - &       - &      - &     822 &   0.38 &   2.18  \\
F20B &   6.05 &     307 &  12.98 &   9.08 &     933 &   0.45 &   2.86 &   0.76 &       - &       - &       - &       - &       - &      - &  1\,148 &   0.14 &   1.11  \\
\hline
H12B &   5.16 &     474 &  17.09 &  25.92 &     632 &   0.27 &   1.05 &   1.46 &       - &       - &       - &       - &       - &      - &     751 &   0.09 &   0.48  \\
\hline
\end{tabular}

\caption{ Radius, $R$, equatorial rotation velocity, $v$, and
equatorial specific angular momentum, $\jeq$ (in units of
$\Ep{18}\,\junit$) at the ZAMS, during the RSG phase (average values),
during the blue loop (average values), and for the final models.
Additionally, maximum values of rotation velocity, $\vmax$, and
specific angular momentum, $\jeqm$, during the blue loop are given.
Dashes indicate the absence of a blue loop.  Lifetime during the MS,
RSG, and BSG phases are respectively given in columns $\tMS$, $\tRSG$,
and $\tBSG$.
\lTab{rot}}
\end{table}
\end{landscape}


\clearpage

\begin{figure}
\epsscale{1.0}
\plotone{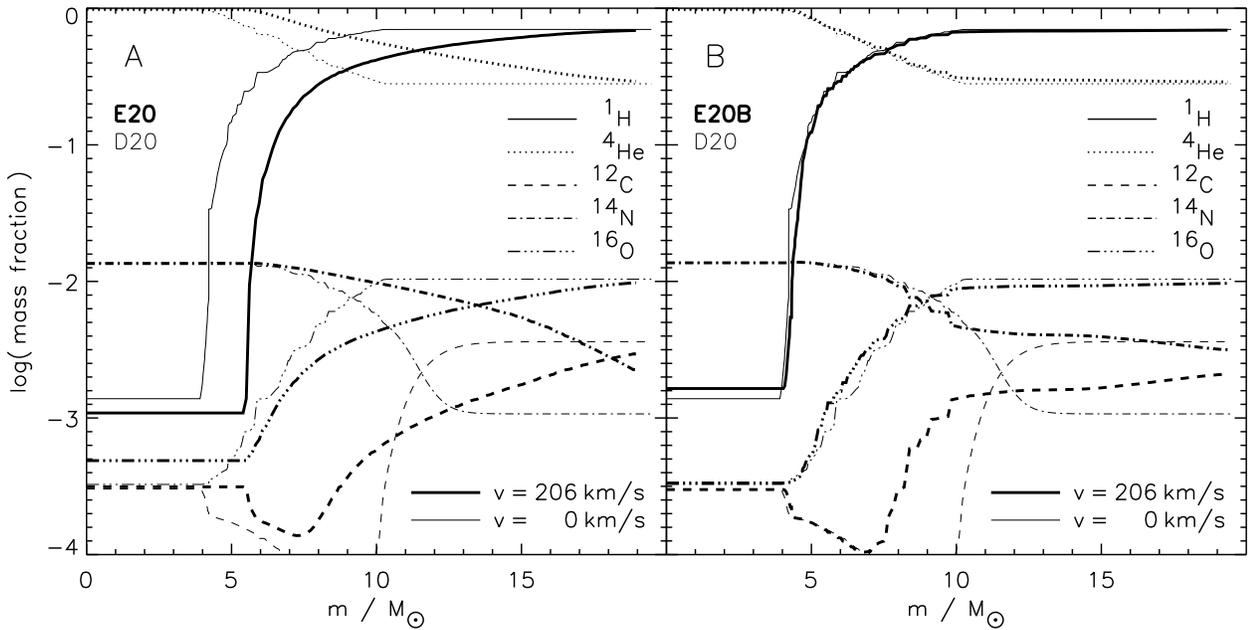}
\caption{ Mass fractions of different isotopes as a function of the
mass coordinate, $m$, at core hydrogen exhaustion.  Compared are the
chemical structures of rotating (thick lines) and a non-rotating (thin
lines; same in both Panels; {Model D20}) $20\,\Msun$ models.  The
rotating models have a ZAMS equatorial rotational velocity of $\sim
200\,\kms$.  \Pan{a}: {Model E20}, where rotationally induced mixing
is {\emph{not}} inhibited by $\mu$-gradients.  \Pan{b}: {Model E20B},
where rotationally induced mixing is (slightly; $\fmu=0.05$) inhibited
by $\mu$-gradients.  See also {\Figs{m-X3-20AB} and
\Figff{m-X2-20AB}}.  \lFig{m-X-20AB}}
\end{figure}

\clearpage

\begin{figure}
\epsscale{1.0}
\plotone{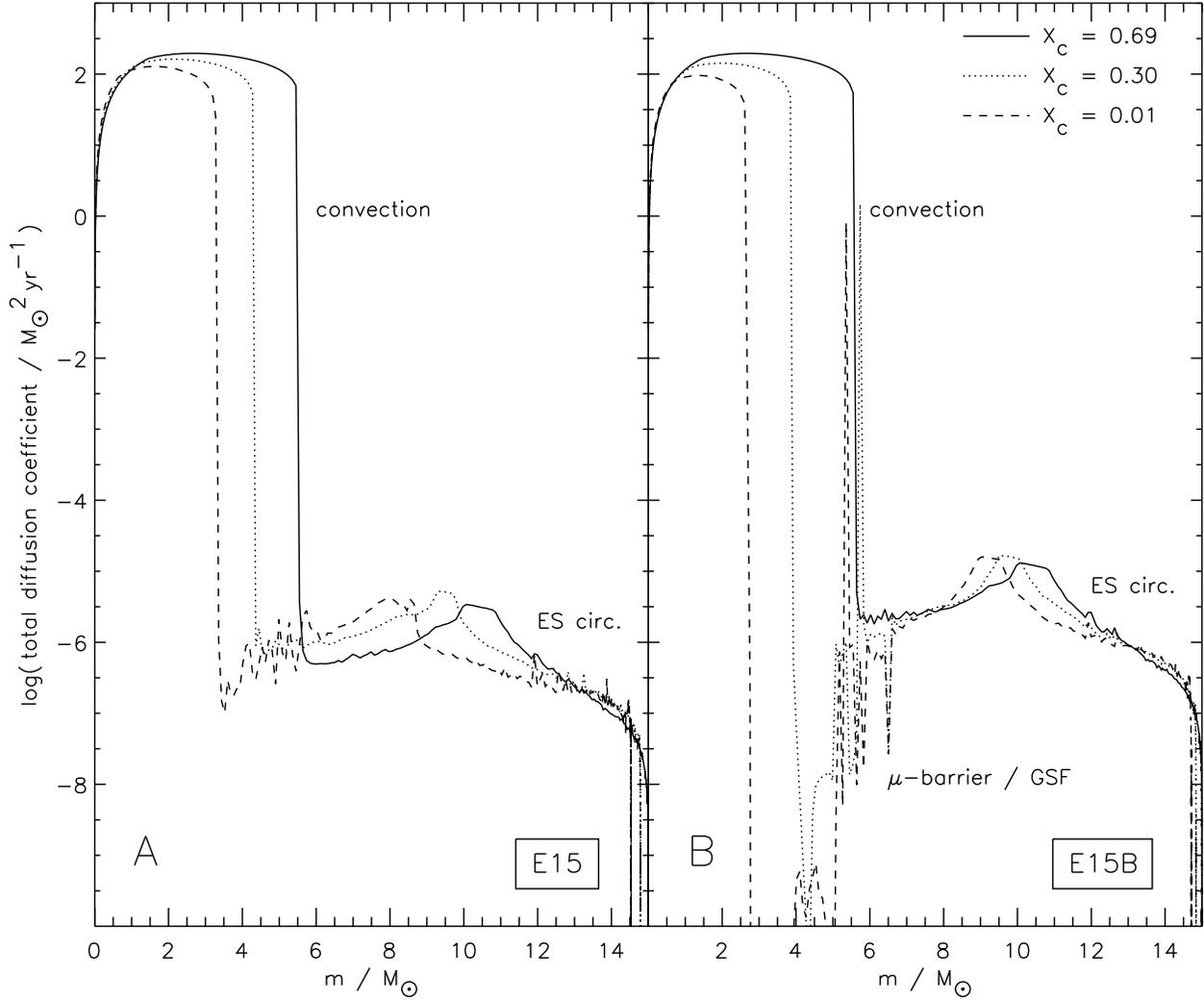}
\caption{Total diffusion coefficient for chemical mixing as a function
of mass coordinate, $m$, at three different stages of central hydrogen
burning: 1\,\% of central hydrogen, by mass, burnt ($\Xc=0.69$; solid
line), and at central hydrogen mass fractions of 30\,\% ($\Xc=0.30$;
dotted line) and 1\,\% ($\Xc=0.01$; dashed line).  \Pan{a}:
Model E15, rotationally induced mixing is {\emph{not}} inhibited by
$\mu$-gradients.  \Pan{b}: Model E15B, rotationally induced mixing
is (slightly; $\fmu=0.05$) inhibited by $\mu$-gradients.  The labels
in the figure indicate the regime of diffusion coefficient for
convection and Eddington-Sweet circulation (ES circ.).  For
Model E15B, the $\mu$-barrier is indecated.  In that region, the
Eddington-Sweet circulation is inhibited and the
Goldreich-Schubert-Fricke instability (GSF) is most efficient process.
Note the region of essentially no mixing above the central convection
zone for the two evolved stages ($\Xc=0.30$, $0.01$) of Model E15B.
\lFig{m-D-E15AB}}
\end{figure}

\clearpage

\begin{figure}
\epsscale{1.0}
\plotone{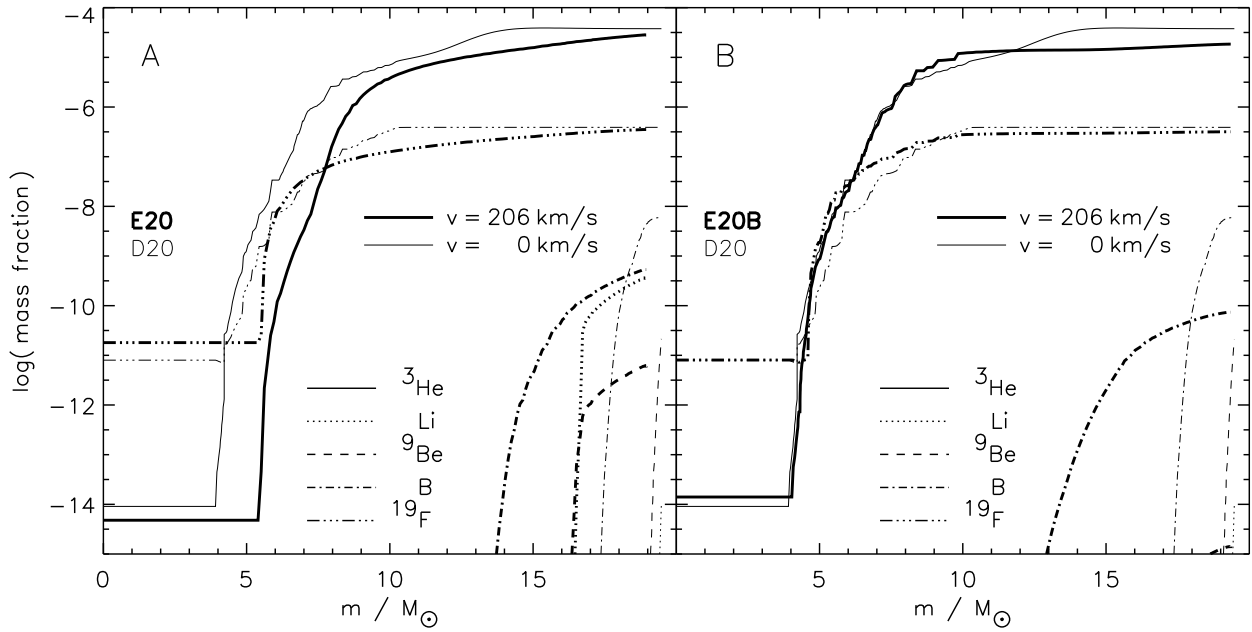}
\caption{ Mass fractions of {\I3{He}} (solid lines), lithium (dotted
lines), {\I9{Be}} (dashed lines), boron (dash-dotted lines), and
{\I{19}F} (dash-triple-dotted lines) as a function of the mass
coordinate, $m$, at core hydrogen exhaustion for the same models as in
{\Fig{m-X-20AB}}.  See also {\Fig{m-X2-20AB}}.  \lFig{m-X3-20AB}}
\end{figure}

\clearpage

\begin{figure}
\epsscale{1.0}
\plotone{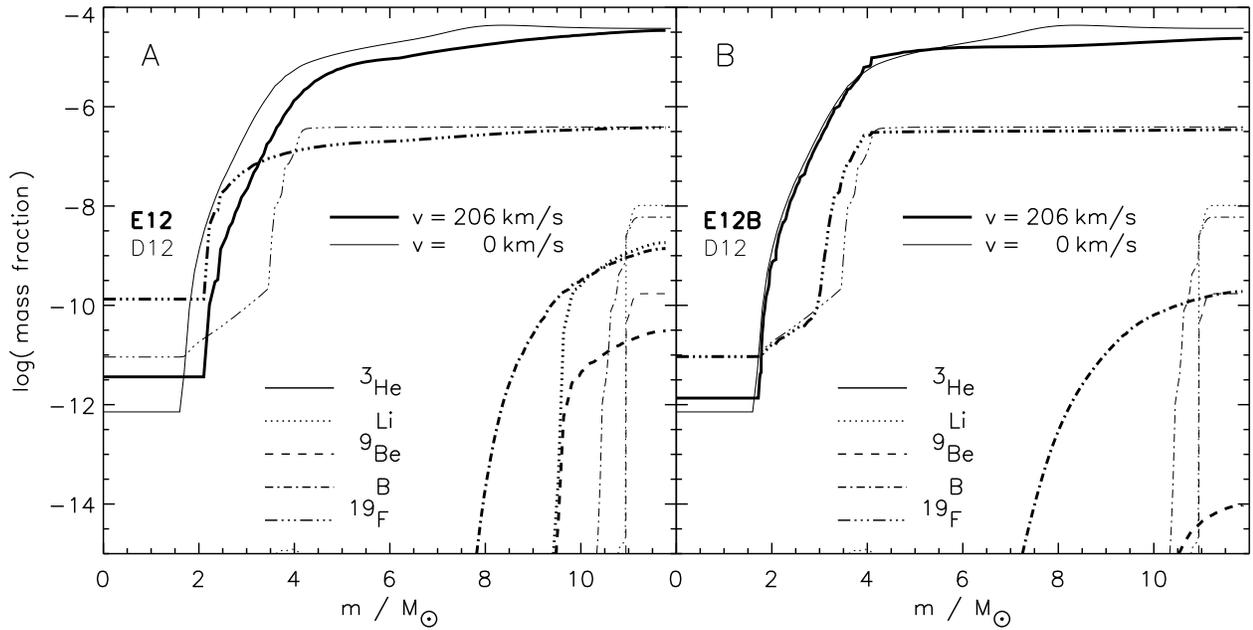}
\caption{ Mass fractions of {\I3{He}} (solid lines), lithium (dotted
lines), {\I9{Be}} (dashed lines), boron (dash-dotted lines), and
{\I{19}F} (dash-triple-dotted lines) as a function of the mass
coordinate, $m$, at core hydrogen exhaustion.  Compared are the
chemical structures of rotating (thick lines) and a non-rotating (thin
lines; same in both Panels; Model D12) $12\,\Msun$ models.  The
rotating models have a ZAMS equatorial rotational velocity of $\sim
200\,\kms$.  \Pan{a}: {Model E12}, where rotationally induced mixing
is {\emph{not}} inhibited by $\mu$-gradients.  \Pan{b}: {Model E12B},
where rotationally induced mixing is (slightly; $\fmu=0.05$) inhibited
by $\mu$-gradients.  See also {\Fig{m-X3-20AB}}.  \lFig{m-X3-12AB}}
\end{figure}

\clearpage

\begin{figure}
\epsscale{1.0}
\plotone{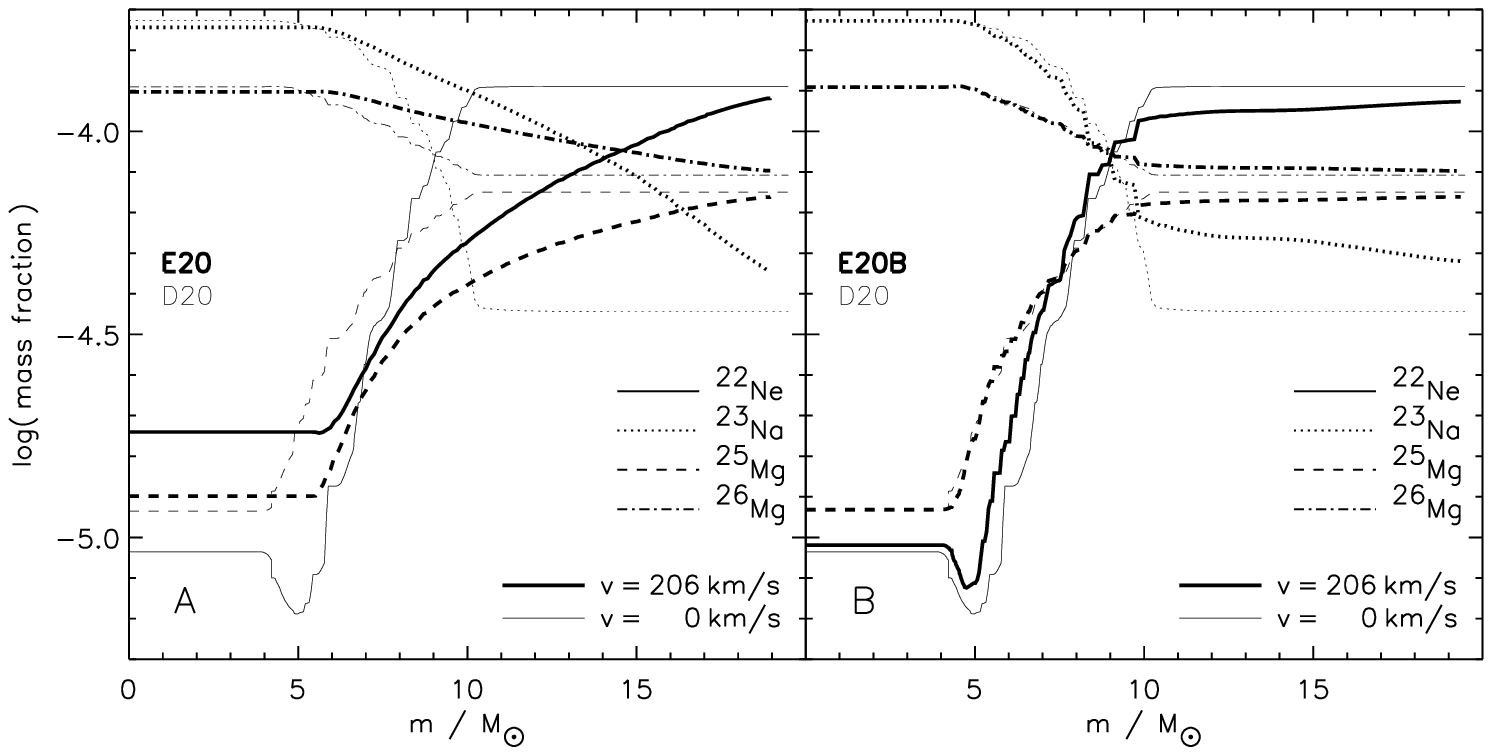}
\caption{
Mass fractions of {\I{22}{Ne}} (solid lines), {\I{23}{Na}} (dotted
lines), {\I{25}{Mg}} (dashed lines), and {\I{26}{Mg}} (dash-dotted
lines) as a function of the mass coordinate, $m$, at core hydrogen
exhaustion for the same models as in {\Fig{m-X-20AB}}.  See also
{\Fig{m-X3-20AB}}.
\lFig{m-X2-20AB}}
\end{figure}

\clearpage

\begin{figure}
\epsscale{1}
\plotone{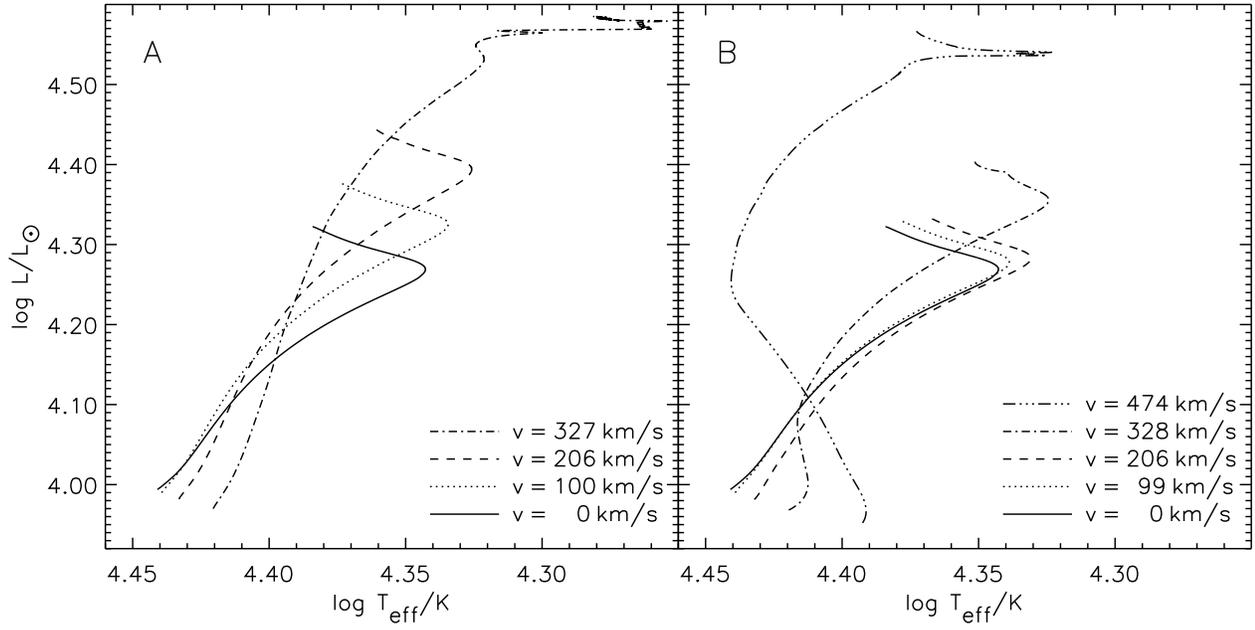}
\caption{
Main sequence evolution of $12\,\Msun$ stars with different ZAMS
rotational velocities in the HR diagram.
{\Pan{a}}: Models D12, G12, E12, and
F12, where rotationally induced mixing is assumed insensitive to
$\mu$-gradients.
{\Pan{b}}: Models D12, G12B, E12B,
F12B, and H12B, where rotationally induced mixing is
assumed to be sensitive to $\mu$-gradients ($\fmu=0.05$).  
\lFig{MS-HRD-12AB}}
\end{figure}

\clearpage

\begin{figure}
\epsscale{1}
\plotone{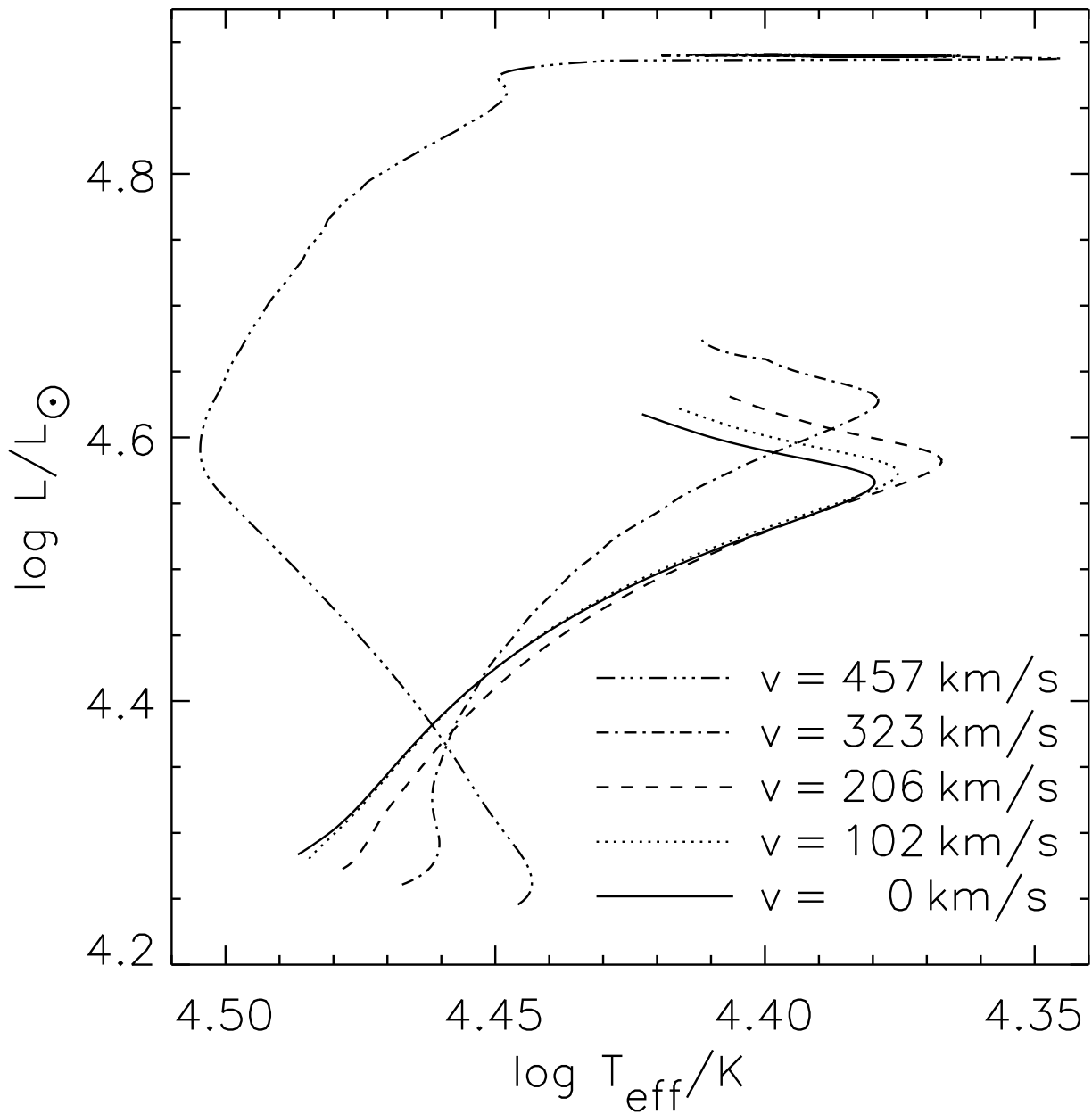}
\caption{
Core hydrogen burning evolution in the HR diagram of $15\,\Msun$
models with ZAMS rotational velocities of $0\,\kms$ ({Model D15};
solid line), $102\,\kms$ ({Model G15B}; dotted line), $206\,\kms$
({Model E15B}; dashed line), $323\,\kms$ ({Model F15B}; dash-dotted
line), and $457\,\kms$ ({Model H15B}; dash-triple-dotted line).
\lFig{MS-HRD-15B}}
\end{figure}

\clearpage

\begin{figure}
\epsscale{1}
\plotone{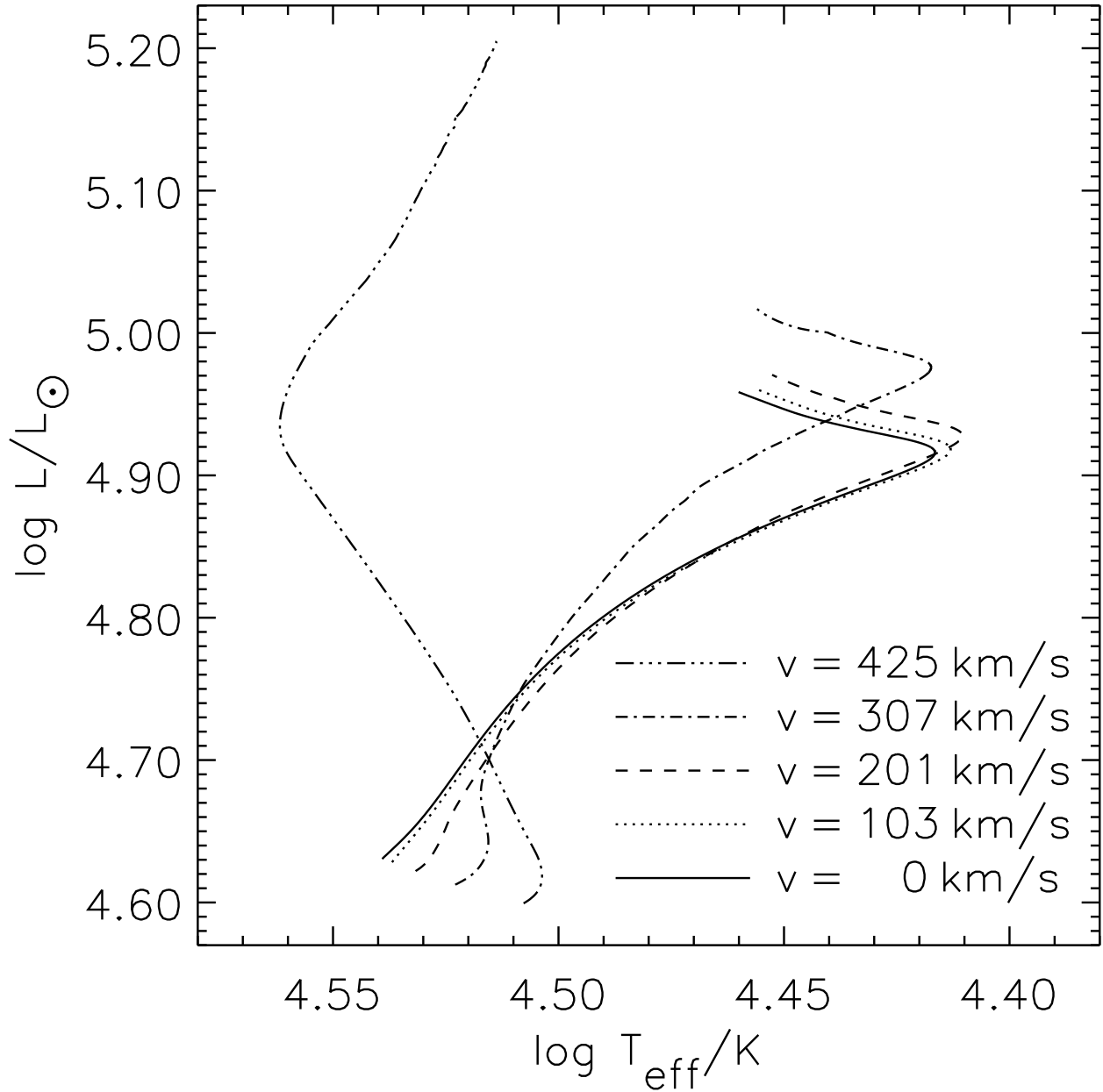}
\caption{
Core hydrogen burning evolution in the HR diagram of $20\,\Msun$
models with ZAMS rotational velocities of $0\,\kms$ ({Model D20};
solid line), $103\,\kms$ ({Model G20B}; dotted line), $201\,\kms$
({Model E20B}; dashed line), $307\,\kms$ ({Model F20B}; dash-dotted
line), and $425\,\kms$ ({Model H20B}; dash-triple-dotted line).  The
fastest rotating star becomes a Wolf-Rayet star at a central hydrogen
abundance of $4\,\%$; the track of this star is discontinued at that
point of evolution.
\lFig{MS-HRD-20B}}
\end{figure}

\clearpage

\begin{figure}
\epsscale{0.8}
\plotone{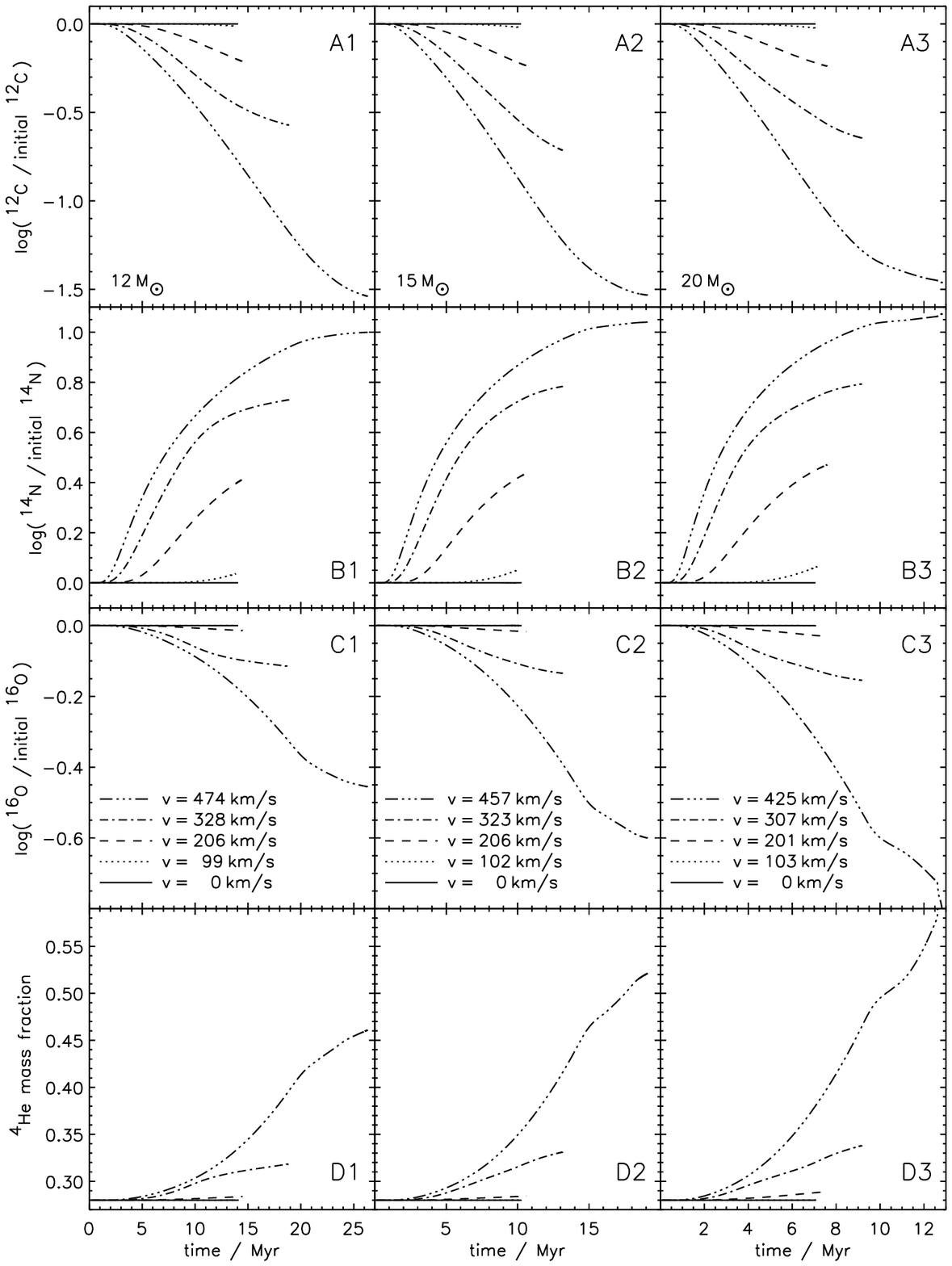}
\caption{ 
Evolution of the surface abundances during central hydrogen burning of
$12\,\Msun$ (left column; Models D12, G12B, E12B, F12B, and H12B),
$15\,\Msun$ (middle column; Models D15, G15B, E15B, F15B, and H15B),
and $20\,\Msun$ (right column; Models D20, G20B, E20B, F20B, and H20B)
stars for different ZAMS equatorial rotational velocities (see legend)
as a function of time.  Displayed are {\I{12}{C}} (\PanRange{a1}{a3}),
{\I{14}{N}} (\PanRange{b1}{b3}), and {\I{16}{O}} (\PanRange{c1}{c3})
relative to their initial abundance, and the {\I{4}{He}} mass fraction
(\PanRange{d1}{d3}).
\lFig{t-CNOHe}}
\end{figure}

\clearpage

\begin{figure}
\epsscale{0.8}
\plotone{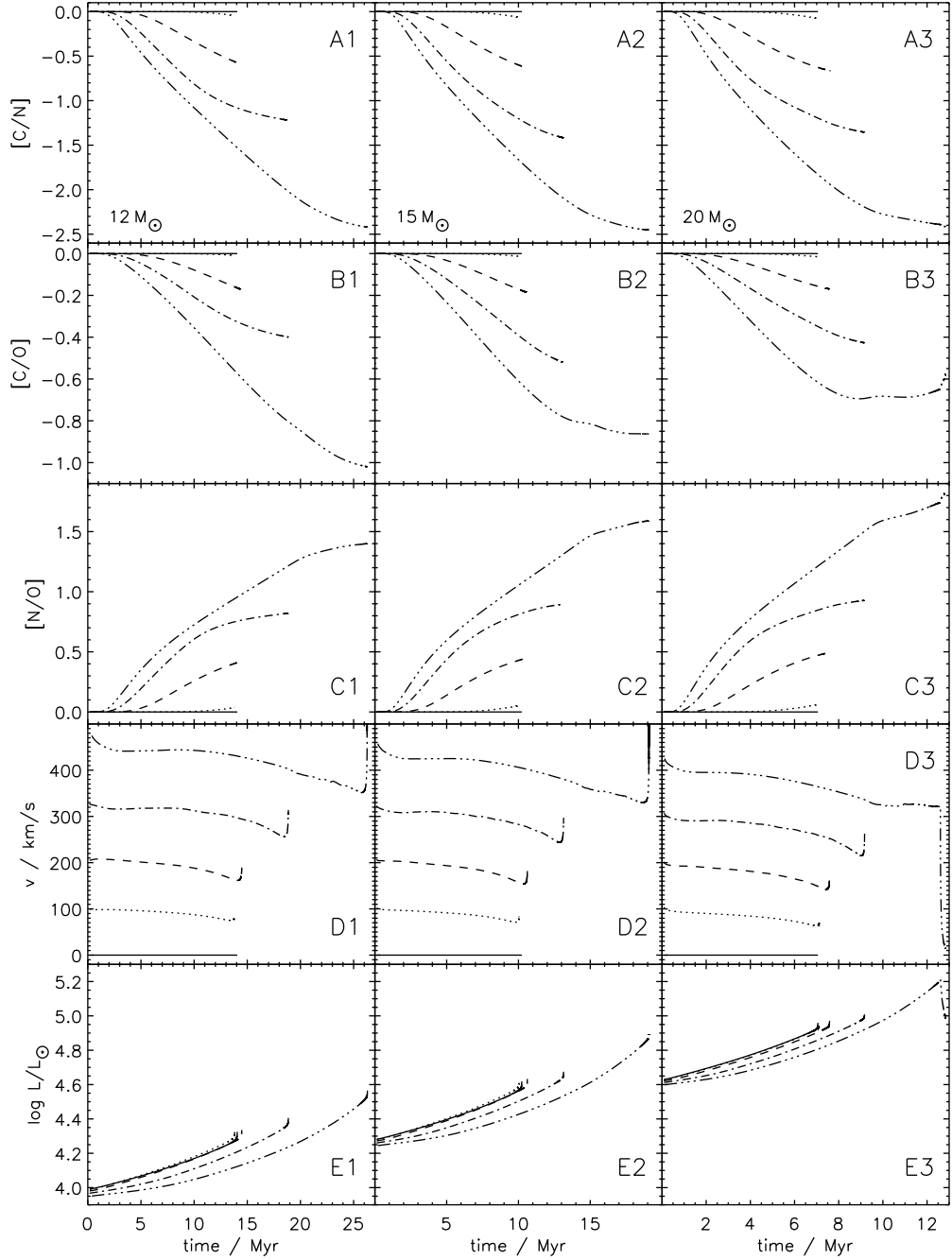}
\caption{
Same models and abscissa as in {\Fig{t-CNOHe}}, but the ratio of
carbon to nitrogen (\PanRange{a1}{a3}), the ratio of carbon to oxygen
(\PanRange{b1}{b3}), and the ratio of nitrogen to oxygen
(\PanRange{c1}{c3}) are displayed, all relative to the abundance
ratios in the initial models.  In {\PanRange{d1}{d3}} the evolution of
the equatorial rotational velocity is displayed and in
{\PanRange{e1}{e3}} that of the luminosity.
\lFig{t-CNOvL}}
\end{figure}

\clearpage

\begin{figure}
\epsscale{0.8}
\plotone{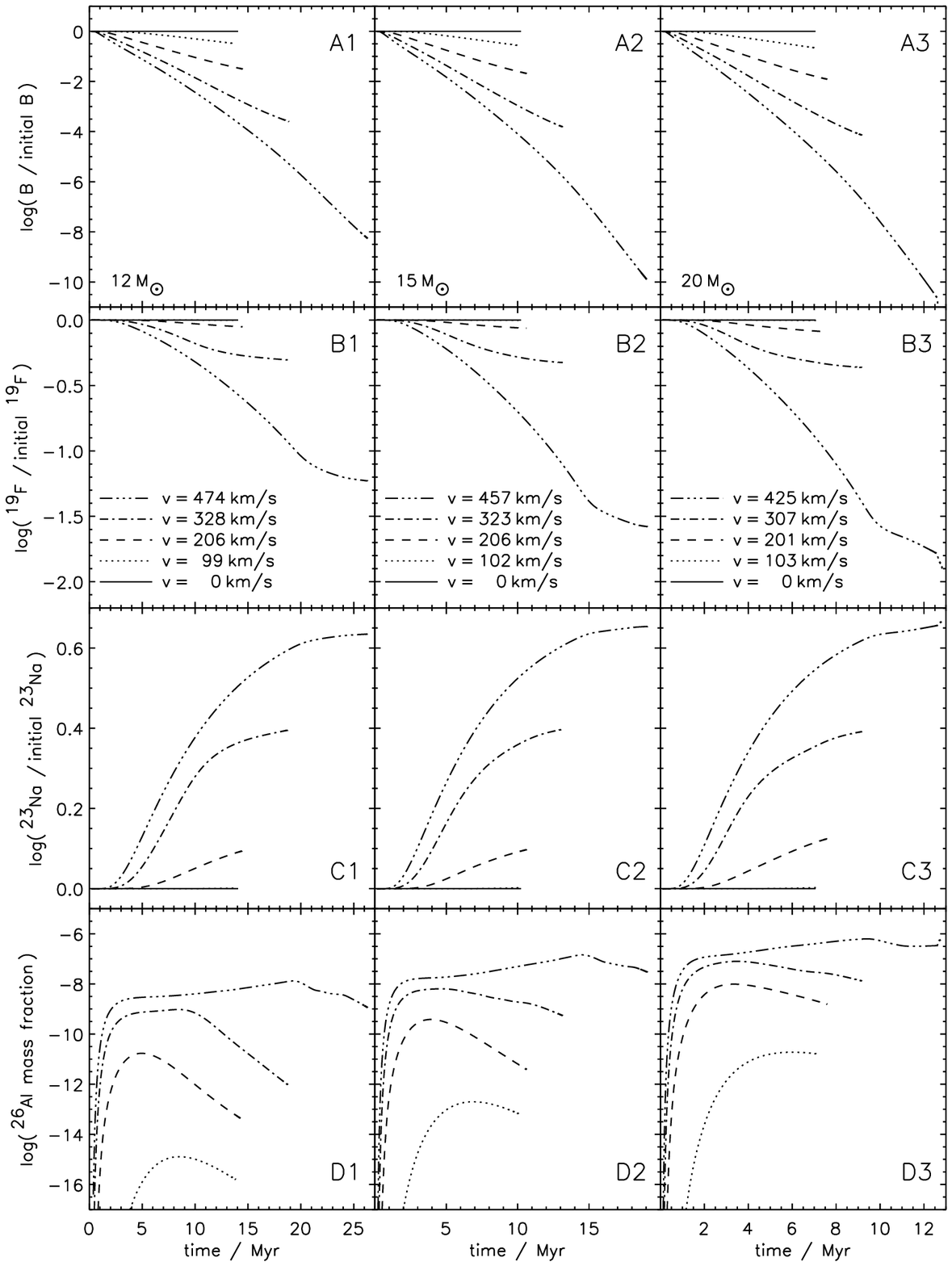}
\caption{ 
Same models and abscissa as in {\Fig{t-CNOHe}}, but the evolution of
the surface abundances of boron (\PanRange{a1}{a3}), {\I{19}{F}}
(\PanRange{b1}{b3}), and {\I{23}{Na}} (\PanRange{c1}{c3}) are given
relative to their initial values; for the radioactive isotope
{\I{26}{Al}} (\PanRange{d1}{d3}) the absolute mass fraction is
displayed.
\lFig{t-BFNaAl}}
\end{figure}

\clearpage

\begin{figure}
\epsscale{1.0}
\plotone{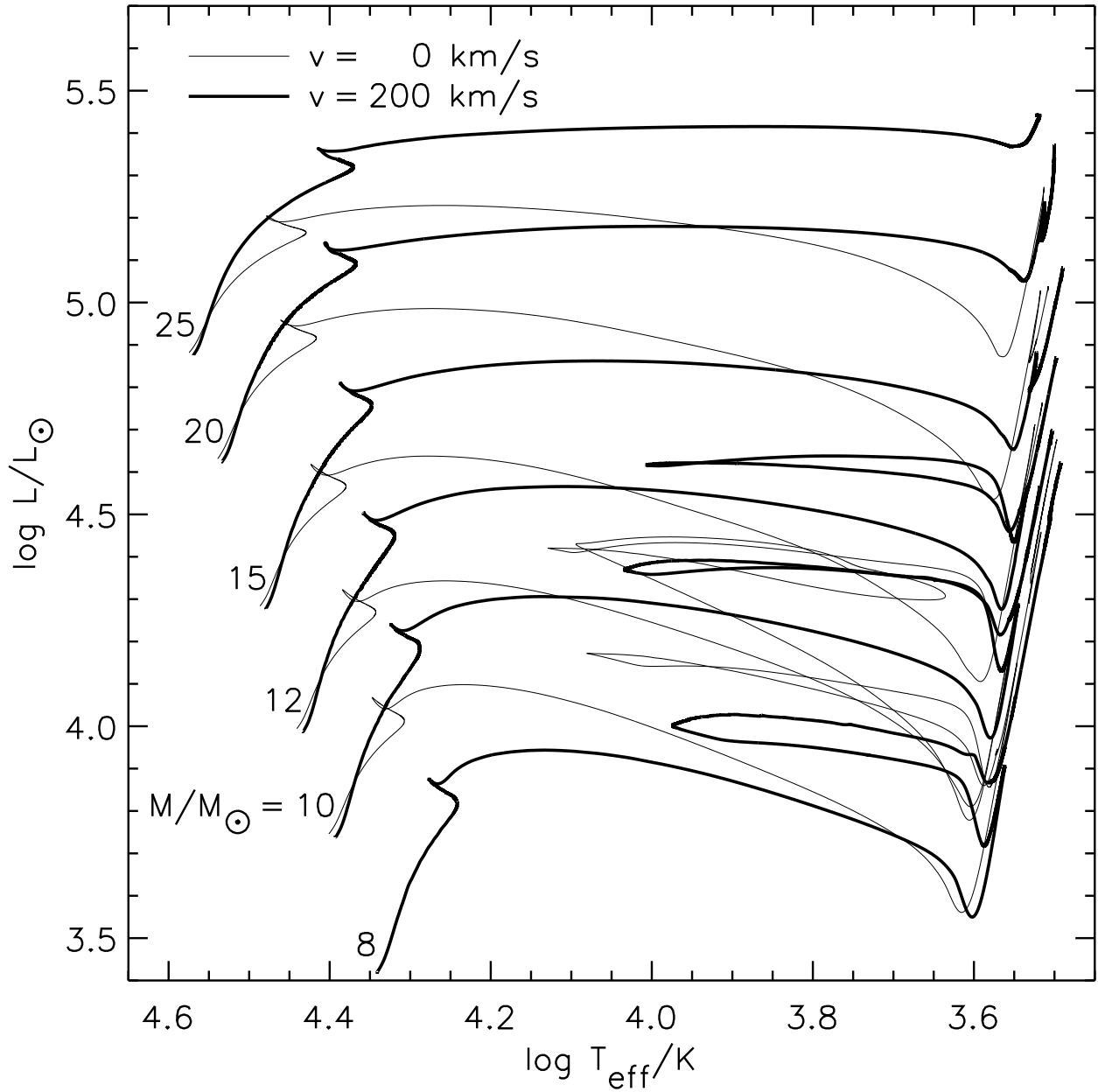}
\caption{
Evolutionary tracks in the HR diagram of rotating models in the mass
range $8\,\Msun$ to $25\,\Msun$ (thick lines) and non-rotating
$10\,\Msun$ to $25\,\Msun$ models (thin lines) from the ZAMS to the
presupernova stage.  The tracks are labeled with the initial masses
(in $\Msun$).  The tracks are for Models D10, D12, D15, D20, and D25 (no
rotation) and for Models E08, E10, E12, E15, E20, E25 (ZAMS
equatorial rotational velocities of $\sim 200\,\kms$).  The track of
the rotating Model E25 is only shown until it evolves into a
Wolf-Rayet star, during core helium burning.  See also {\cite{HLW98}}.
\lFig{HRD-DE}}
\end{figure}

\clearpage

\begin{figure}
\epsscale{1.0}
\plotone{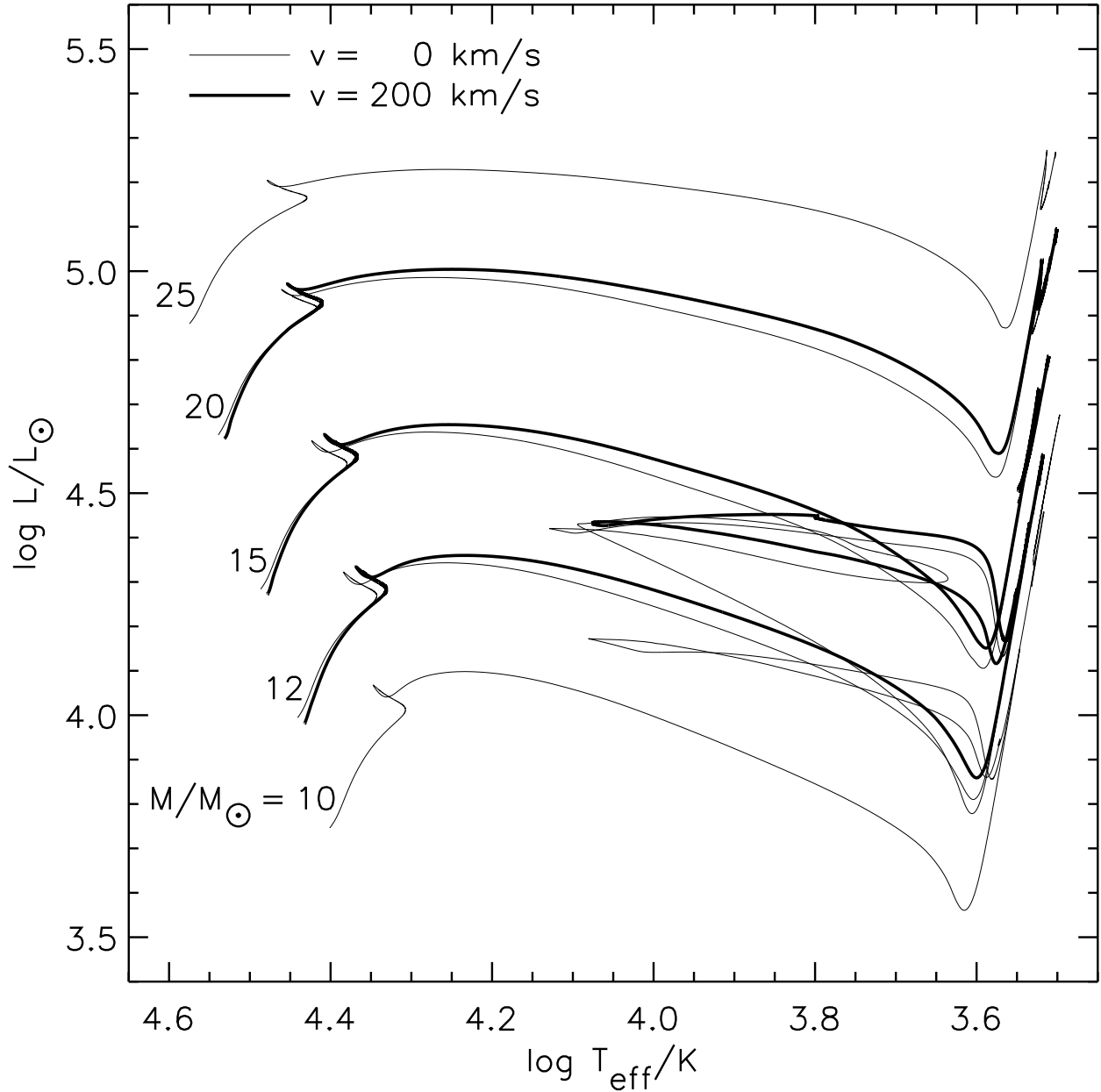}
\caption{Evolutionary tracks in the HR diagram of rotating models in
the mass range $12\,\Msun$ to $20\,\Msun$ (thick lines) and of
non-rotating models in the mass range $10\,\Msun$ to $25\,\Msun$ (thin
lines) from core hydrogen ignition to the presupernova stage.  The
tracks are labeled with the initial masses (in $\Msun$).  The tracks
are for Models D10, D12, D15, D20, and D25 (non-rotating) and for
Models E12B, E15B, and E20B (equatorial rotational velocities of
$\sim 200\,\kms$).  See also {\Fig{HRD-DE}}.
\lFig{HRD-DEB}}
\end{figure}

\clearpage

\begin{figure}
\epsscale{1.0}
\plotone{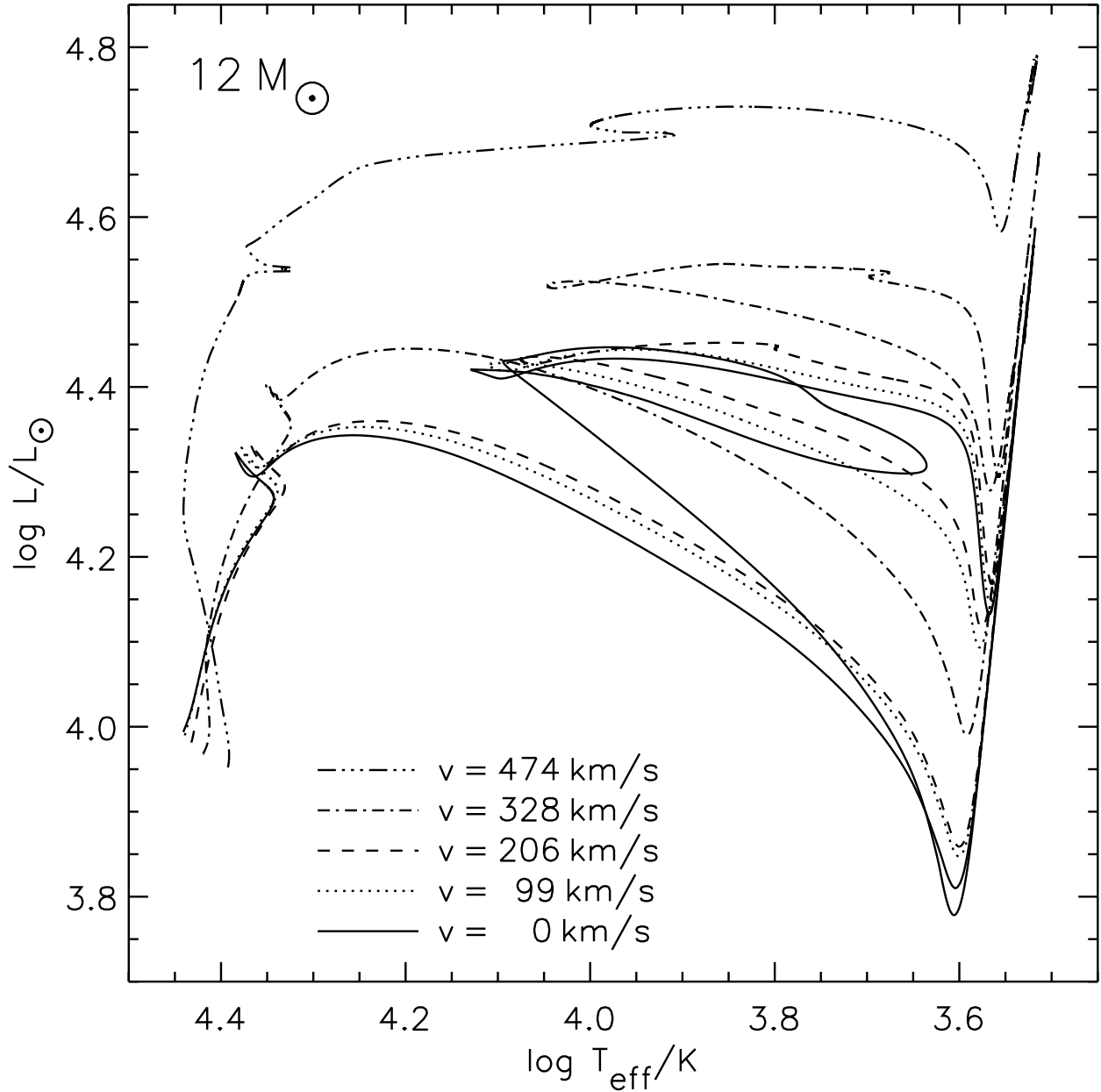}
\caption{
Evolutionary tracks in the HR diagram of stars with an initial mass of
$12\,\Msun$ and ZAMS equatorial rotational velocities of $0\,\kms$
(Model D12, solid line), $99\,\kms$ (Model G12B, dotted line),
$206\,\kms$ (Model E12B, dashed line), $328\,\kms$ (Model F12B,
dash-dotted line), and $474\,\kms$ (Model H12B, dash-triple-dotted
line).  The evolution during central hydrogen burning is magnified in
{\FIG{b}{MS-HRD-12AB}}.  The tracks start at the ZAMS and are followed
until the phase listed in \cite[][]{HLW00:I}.
\lFig{HRD-12B}}
\end{figure}

\clearpage

\begin{figure}
\epsscale{1.0}
\plotone{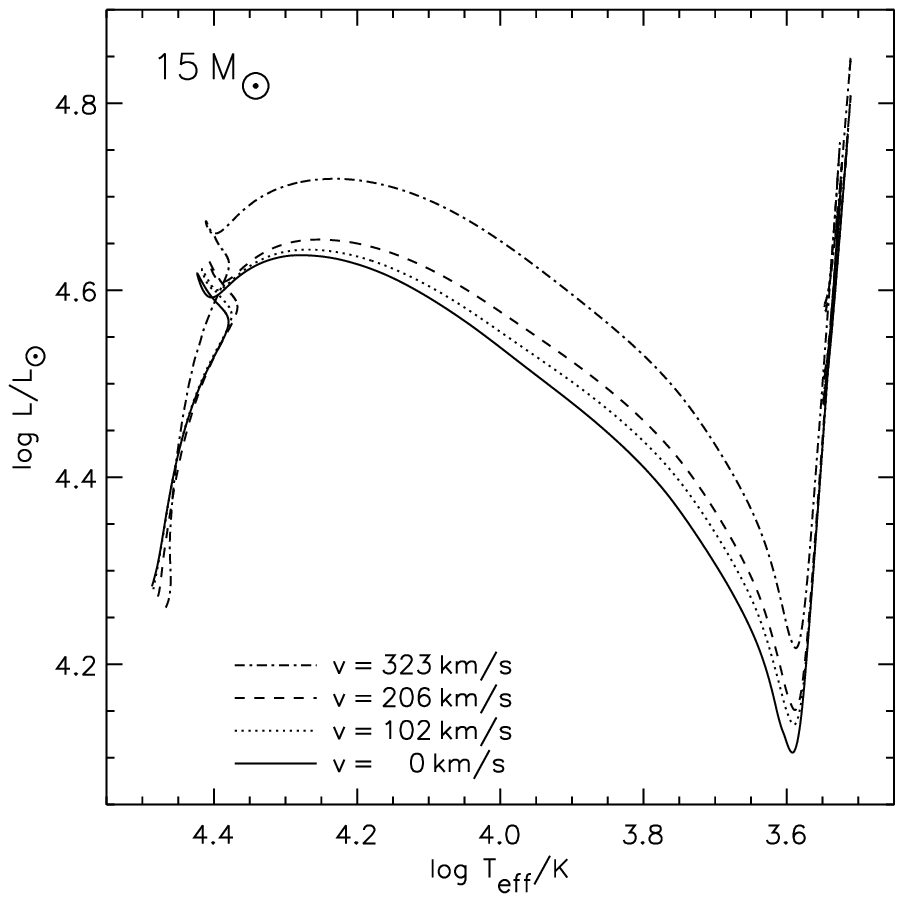}
\caption{
Evolutionary tracks in the HR diagram of stars with an initial mass of
$15\,\Msun$ and ZAMS equatorial rotational velocities of $0\,\kms$
(Model D15, solid line), $102\,\kms$ (Model G15B, dotted line),
$206\,\kms$ (Model E15B, dashed line), and $323\,\kms$ (Model F15B,
dash-dotted line) up to core collapse.  The evolution during central
hydrogen burning is magnified in {\Fig{MS-HRD-15B}}.
\lFig{HRD-15B}}
\end{figure}

\clearpage

\begin{figure}
\epsscale{1.0}
\plotone{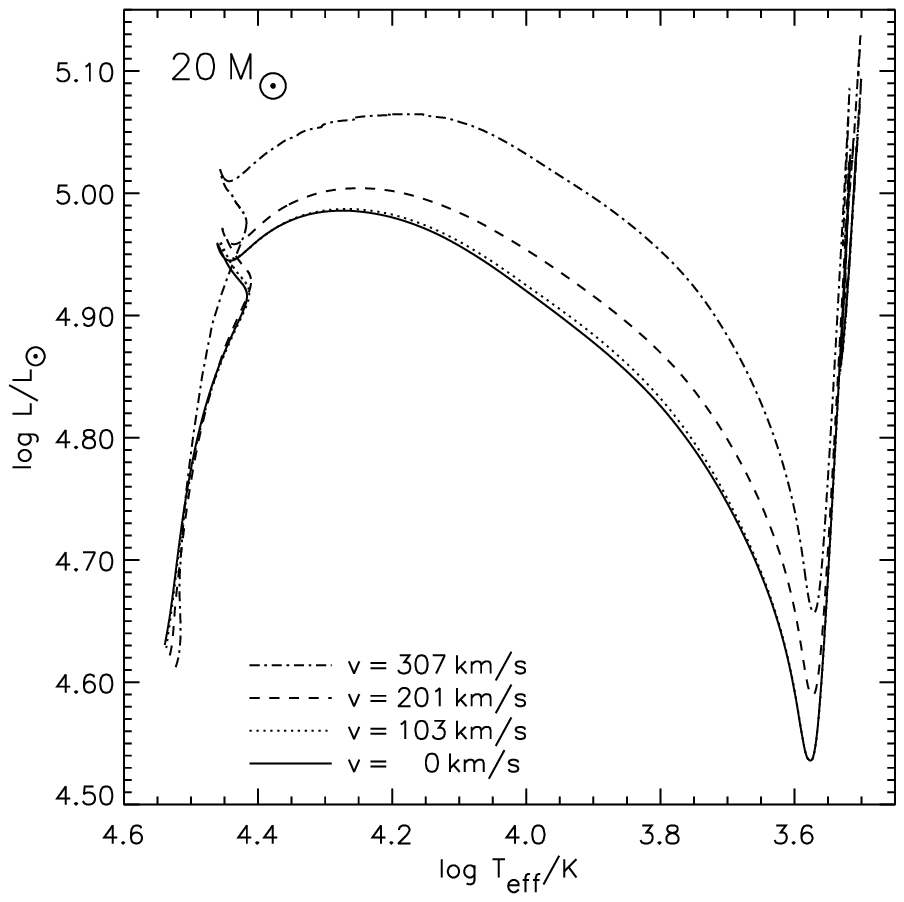}
\caption{
Evolutionary tracks in the HR diagram of stars with an initial mass of
$15\,\Msun$ and ZAMS equatorial rotational velocities of $0\,\kms$
(Model D20, solid line), $103\,\kms$ (Model G20B, dotted line),
$201\,\kms$ (Model E20B, dashed line), and $307\,\kms$ (Model F20B,
dash-dotted line) up to core collapse.  The evolution during central hydrogen burning is
magnified in {\Fig{MS-HRD-20B}}.
\lFig{HRD-20B}}
\end{figure}

\clearpage

\begin{figure}
\epsscale{0.4}
\plotone{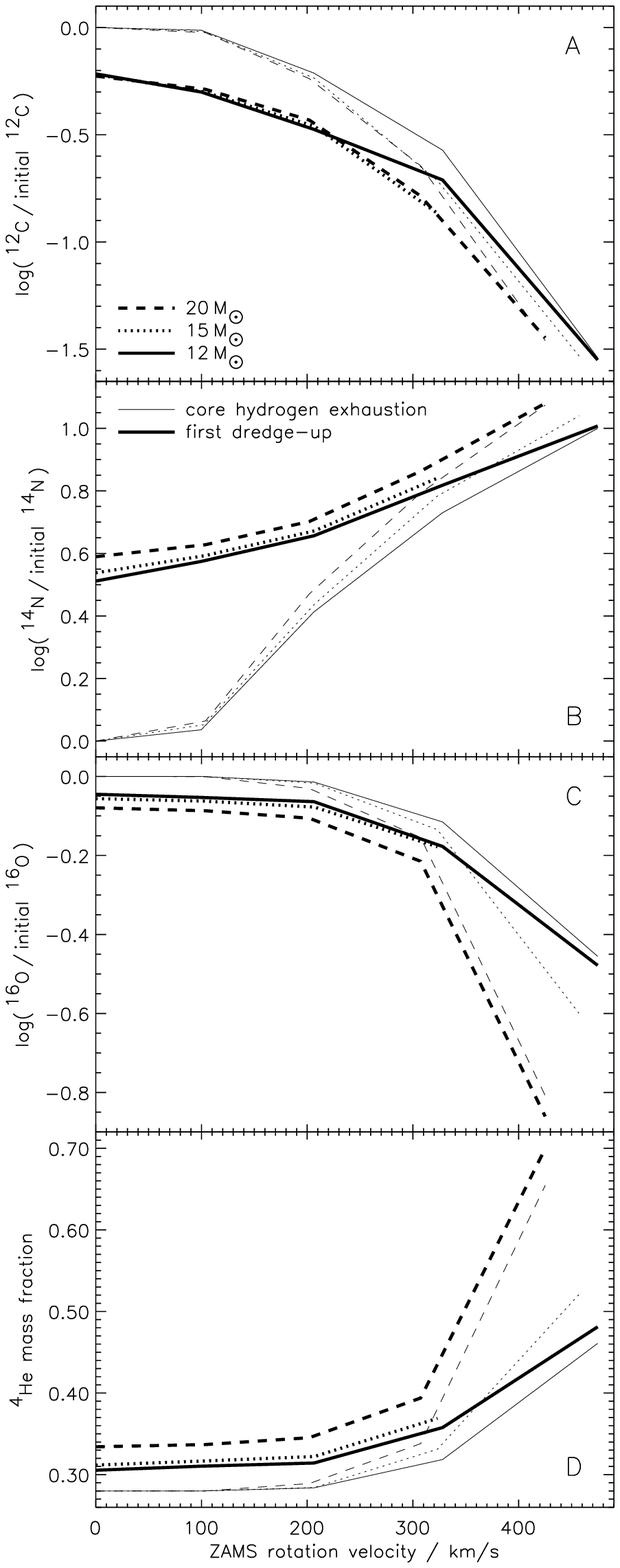}
\caption{ 
Surface mass fractions of {\I{12}{C}} (\Pan{a}), {\I{14}{N}}
(\Pan{b}), and {\I{16}{O}} (\Pan{c}), relative to the initial values,
as a function the initial equatorial rotation rate.  {\Pan{d}} gives
the {\I{4}{He}} mass fraction.  Solid, dotted and dashed lines give
the values for the $12\,\Msun$ (Models D12, G12B, E12B, F12B, and
H12B), $15\,\Msun$ (Models D15, G15B, E15B, F15B, and H12B), and
$20\,\Msun$ (Models D20, G20B, E20B, F20B, and H20B) stars,
respectively.  The thin lines show the surface composition at the end
of central hydrogen burning (\Fig{t-CNOHe}), the thick lines those
after the first dredge-up, i.e., when the star has first become a red
supergiant after termination of central hydrogen burning.  For the
fastest rotating $15\,\Msun$ star we have no post-dredge-up data.
\lFig{v-CNOHe-RSG}}
\end{figure}

\clearpage

\begin{figure}
\epsscale{0.5}
\plotone{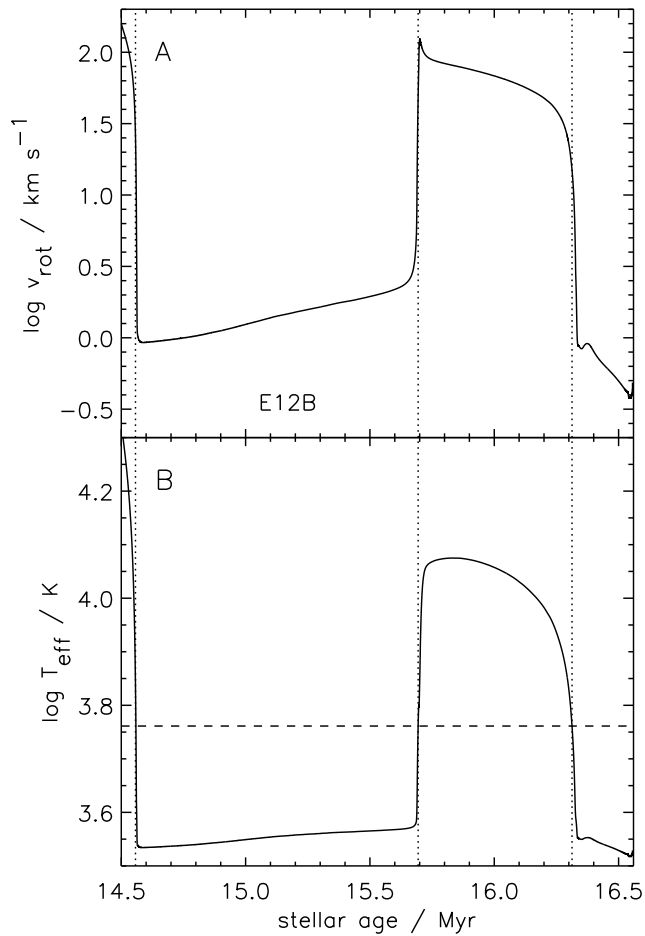}
\caption{ 
Equatorial surface rotation velocity (\Pan{a}) and effective
temperature (\Pan{b}) for Model E12B as a function of time, displayed
from the end of central hydrogen burning until core collapse.  The
dashed line indicates the effective temperature of the sun, and dotted
lines separate ``blue'' from ``red'' phases.  Between an age of $15.7$
and $16.3\,\Myr$ the star undergoes a blue loop, while it spends the
major part of central helium burning and the time from central helium
depletion till core collapse as a RSG.  Note that the maximum and
minimum rotational velocity during the blue loop deviate by about one
order of magnitude and that the rotational velocity after the blue loop
is notably slower than before the loop.
\lFig{t-vTeff-E12B}}
\end{figure}

\end{document}